%% file: 0_main.tex
\documentclass[journal]{IEEEtran}

\usepackage{amsmath}
\usepackage{amsbsy}
\usepackage{amssymb}
\usepackage{comment}
\usepackage[utf8]{inputenc}
\usepackage{graphicx}
\graphicspath{{Images/}}
\usepackage[ampersand]{easylist}
\usepackage[noadjust]{cite}
\usepackage{xcolor}
\usepackage{float}
\usepackage{booktabs}
\usepackage[flushleft]{threeparttable}
\usepackage{adjustbox}
\usepackage[hidelinks]{hyperref}
\usepackage{siunitx}
\usepackage{wasysym}

\newcommand{\lidar}{lidar}
\newcommand{\Lidar}{Lidar}

\title{
Dense 3D Reconstruction Through \Lidar{}: \\A Comparative Study on Ex-vivo Porcine Tissue}

\author{Guido~Caccianiga$^{1,2,3}$,~\IEEEmembership{Graduate Student Member,~IEEE},
        Julian~Nubert$^{2,3}$,~\IEEEmembership{Graduate Student Member,~IEEE},\\
        Marco~Hutter$^{2,3}$,~\IEEEmembership{Member,~IEEE},
        Katherine~J.~Kuchenbecker$^{1,3}$,~\IEEEmembership{Fellow,~IEEE}%

\thanks{$^{1}$ G. Caccianiga, J. Nubert, and K. J. Kuchenbecker are with the Haptic Intelligence Department, Max Planck Institute for Intelligent Systems, Stuttgart, Germany.}
\thanks{$^{2}$ G. Caccianiga, J. Nubert, and M. Hutter are with the Robotic Systems Lab, ETH Zurich, Switzerland.}
\thanks{Corresponding author: Guido Caccianiga, \href{mailto:caccianiga@is.mpg.de}{\texttt{caccianiga@is.mpg.de}}}

}

\IEEEoverridecommandlockouts
\usepackage{fancyhdr}
\newcommand{\mytitle}{\textbf{Preprint version.} This work has been submitted to the IEEE Transactions on Medical Robotics and Bionics for possible publication. Copyright may be transferred without notice, after which this version may no longer be accessible.} 
\begin{document}

\maketitle
\thispagestyle{fancy}
\IEEEpeerreviewmaketitle

\begin{abstract}
New sensing technologies and more advanced processing algorithms are transforming computer-integrated surgery. While researchers are actively investigating depth sensing and 3D reconstruction for vision-based surgical assistance, it remains difficult to achieve real-time, accurate, and robust 3D representations of the abdominal cavity for minimally invasive surgery. Thus, this work uses quantitative testing on fresh ex-vivo porcine tissue to thoroughly characterize the quality with which a 3D laser-based time-of-flight sensor (\lidar{}) can perform anatomical surface reconstruction. Ground-truth surface shapes are captured with a commercial laser scanner, and the resulting signed error fields are analyzed using rigorous statistical tools. When compared to modern learning-based stereo matching from endoscopic images, time-of-flight sensing demonstrates higher precision, lower processing delay, higher frame rate, and superior robustness against sensor distance and poor illumination. Furthermore, we report on the potential negative effect of near-infrared light penetration on the accuracy of \lidar{} measurements across different tissue samples, identifying a significant measured depth offset for muscle in contrast to fat and liver. Our findings highlight the potential of \lidar{} for intraoperative 3D perception and point toward new methods that combine complementary time-of-flight and spectral imaging.
\end{abstract}


\section{Introduction}\label{sec:intro}
\input{1_Introduction}
\section{Related Work}\label{sec:related_work}
\input{2_RelatedWork}

\section{Rationale}\label{sec:rationale}
\input{3_Rationale}
\section{Setup and Data Acquisition}\label{sec:method}
\input{4.1_Materials}
\section{3D Data Processing and Evaluation}\label{sec:processing}
\input{4.2_Processing}

\section{Experiments}\label{sec:experiments}
\input{5_Experiments}
\section{Results}\label{sec:results}
\input{7_Results}

\section{Discussion}\label{sec:discussion}
\input{8_Discussion}
\section{Conclusions and Future Work}\label{sec:conclusion}
\input{9_Conclusions}

\section*{Acknowledgments}
The authors thank Devin Sheehan for support with the tissue samples, Bernard Javot for help with the experimental setup, and Haliza Mat Husin and Mayumi Mohan for guidance throughout the statistical analysis. The authors thank the International Max Planck Research School for Intelligent Systems (IMPRS-IS) for supporting Guido Caccianiga and the Max Planck ETH Center for Learning Systems (CLS) for supporting Guido Caccianiga and Julian Nubert.

\ifCLASSOPTIONcaptionsoff
  \newpage
\fi
\bibliographystyle{IEEEtran}
\bibliography{IEEEabrv, references_T-MRB_02_23}

\begin{IEEEbiography}[{\includegraphics[trim={20 0 20 0},width=1in,height=1.25in,clip,keepaspectratio]{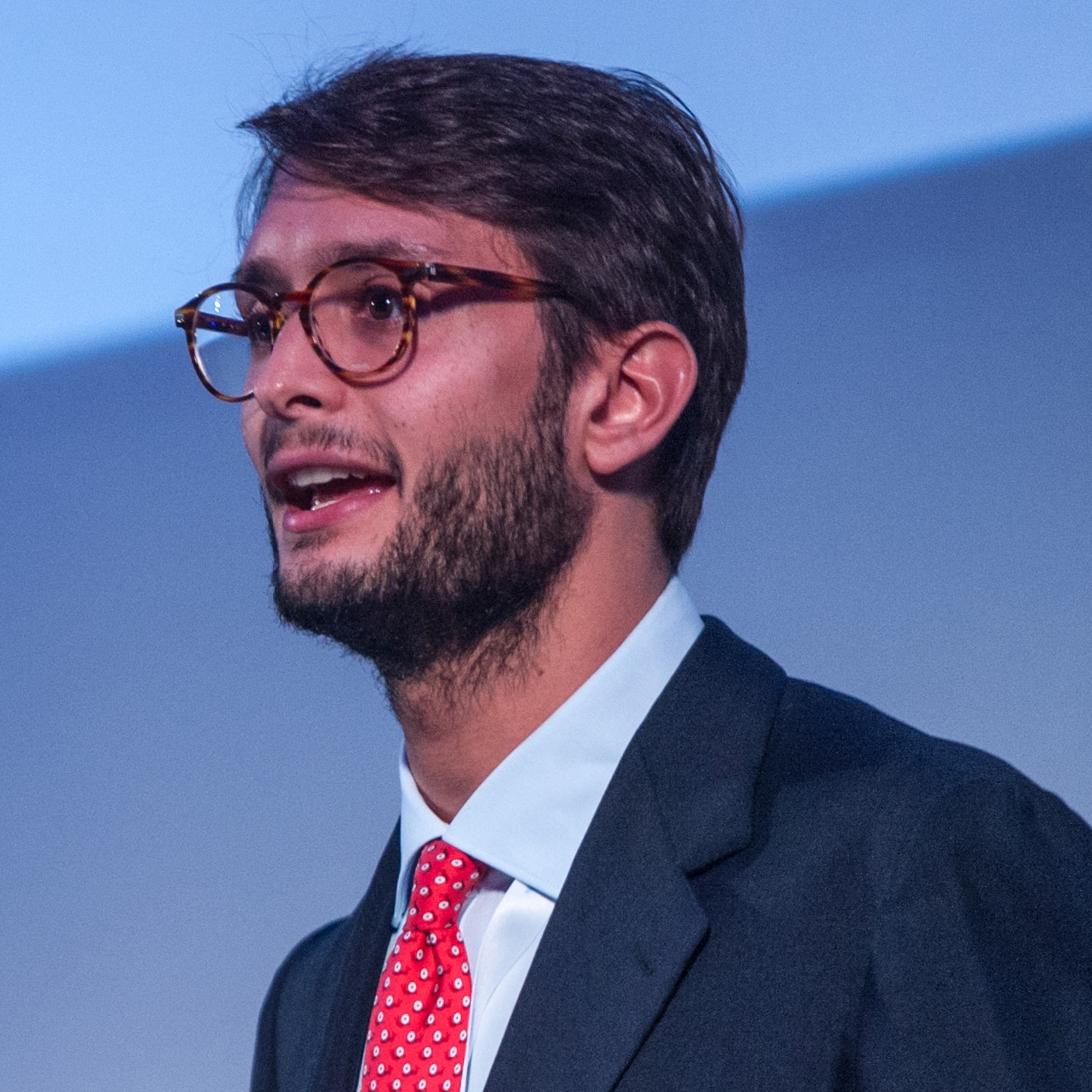}}]{Guido Caccianiga}
received his B.Sc. in Mechanical Engineering in 2016 and M.Sc. in Biomedical Engineering in 2019 both at Politecnico di Milano. He was a visiting scholar at the Laboratory for Computational Sensing and Robotics (LCSR) at Johns Hopkins University (Baltimore, Maryland) and then a research fellow at the Electronic Information and Bioengineering Department (DEIB) of Politecnico di Milano. As of 2020, he is a Computer Science doctoral student in the Haptic Intelligence Department in the Max Planck Institute for Intelligent Systems, where he is a scholar in the International Max Planck Research School for Intelligent Systems (IMPRS-IS) and an affiliated member of the Max Planck ETH Center for Learning Systems. His research focuses on the integration of hardware and software solutions for robot-assisted surgery with specific mention of 3D reconstruction, haptics, AR/VR, and user interface technologies.
\end{IEEEbiography}

\begin{IEEEbiography}[{\includegraphics[trim={400 180 400 0},clip,width=1in,height=1.25in,keepaspectratio]{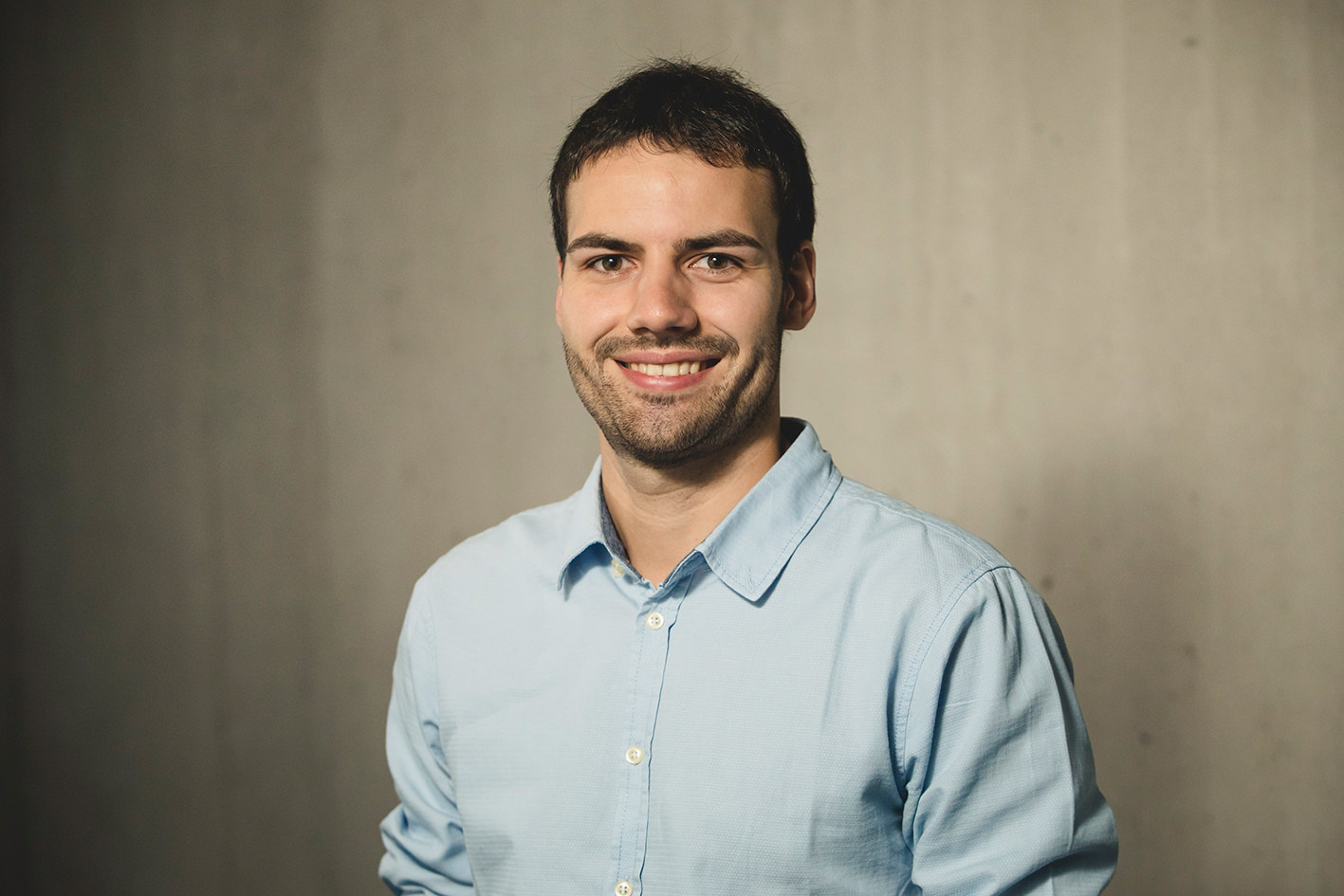}}]%
{Julian Nubert} is a doctoral student in the Max Planck ETH Center for Learning Systems, and part of the Robotic Systems Lab at ETH Zurich and Haptic Intelligence Department at the Max Planck Institute for Intelligent Systems. He received his M.Sc. in Robotics, Systems \& Control in 2020 from ETH Zurich. His research interests lie in the field of robust robot perception, and how it can be used for the deployment of complex robotic systems. Julian received the ETH Silver Medal and the Willi Studer Prize for his accomplishments during his master's studies.
\end{IEEEbiography}

\begin{IEEEbiography}[{\includegraphics[trim={400 180 400 0}, clip,width=1in,height=1.25in,keepaspectratio]{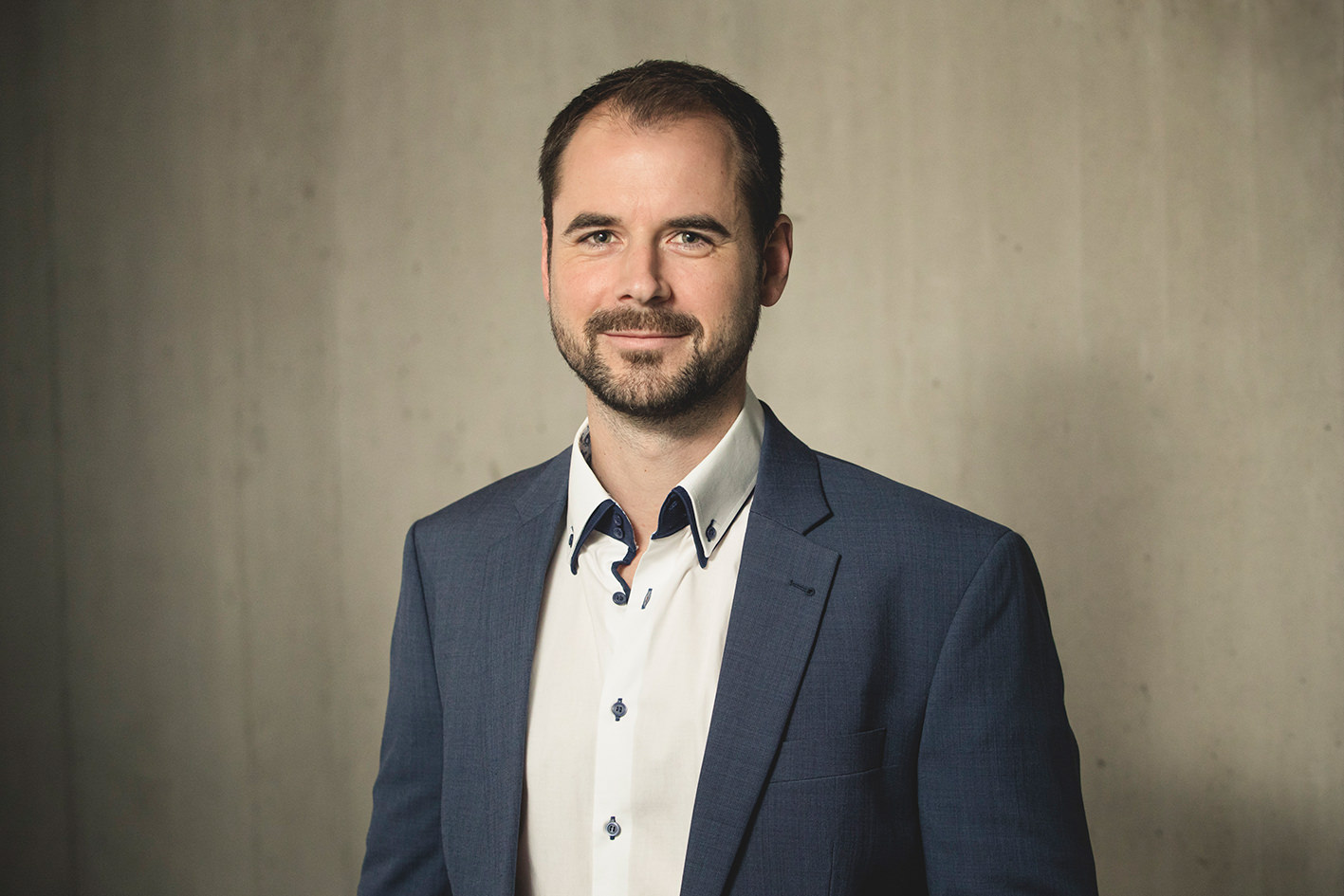}}]%
{Marco Hutter} is Associate Professor for Robotic Systems at ETH Zurich. He received his M.Sc. and Ph.D. from ETH Zurich in 2009 and 2013 in the field of design, actuation, and control of legged robots. His research interests are in the development of novel machines and actuation concepts together with the underlying control, planning, and machine learning algorithms for locomotion and manipulation. Marco is the recipient of an ERC Starting Grant, PI in NCCR Robotics and NCCR Digital Fabrication, PI in various EU projects and international challenges, and co-founder of several ETH startups including ANYbotics AG and Gravis Robotics AG.
\end{IEEEbiography}

\begin{IEEEbiography}[{\includegraphics[trim={120 0 150 0},width=1in,height=1.25in,clip,keepaspectratio]{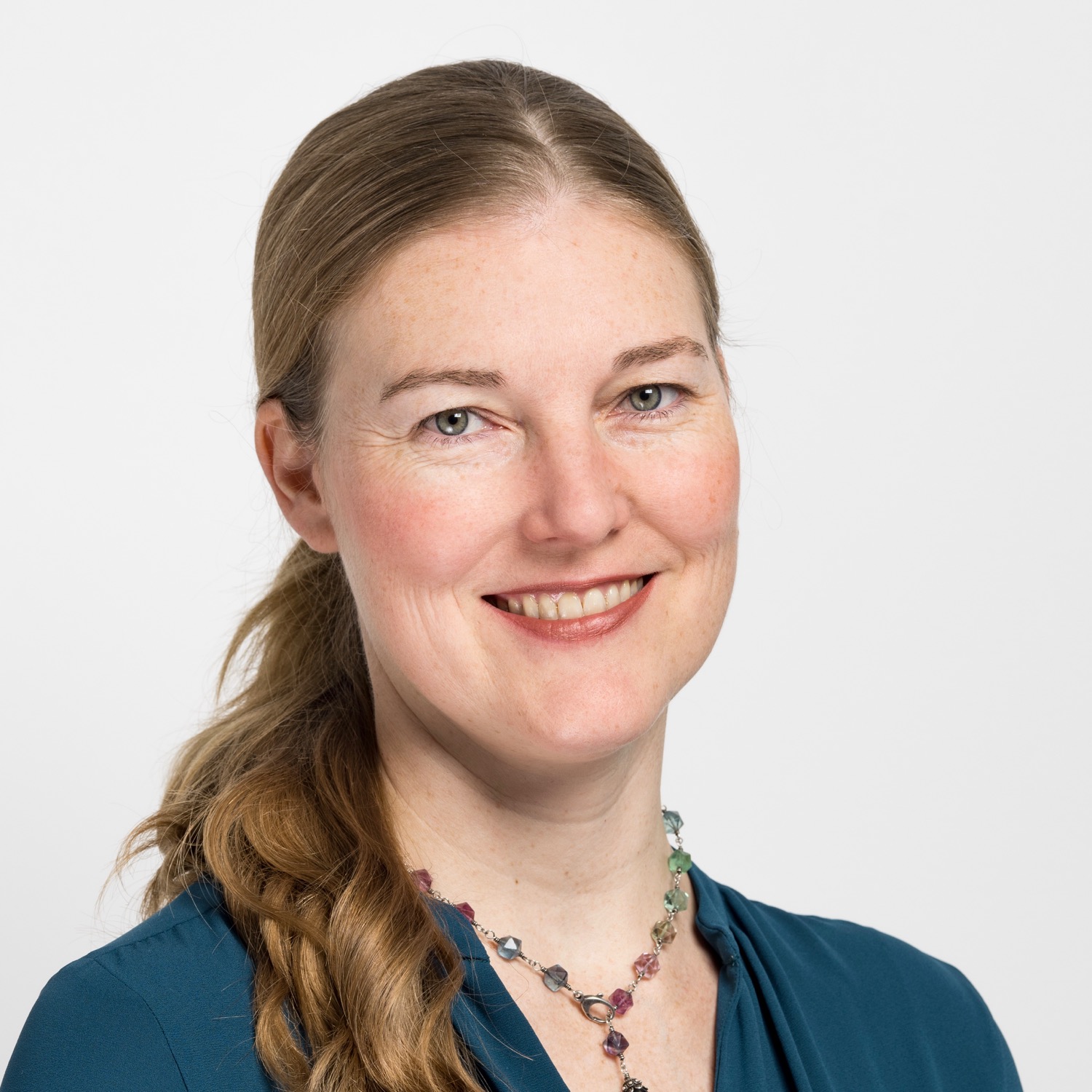}}]{Katherine J. Kuchenbecker} earned B.S., M.S., and Ph.D. degrees in Mechanical Engineering from Stanford University, USA, in 2000, 2002, and 2006, respectively. She was a postdoctoral researcher at the Johns Hopkins University and an Assistant and Associate Professor in the GRASP Lab at the University of Pennsylvania from 2007 to 2016. Since 2017, she has been a Director at the Max Planck Institute for Intelligent Systems in Stuttgart, Germany. Her research blends robotics and human-computer interaction, including work in haptics, teleoperation, physical human-robot interaction, tactile sensing, and medical applications. She has been honored with a 2009 NSF CAREER Award, the 2012 IEEE RAS Academic Early Career Award, a 2014 Penn Lindback Award for Distinguished Teaching, and elevation to IEEE Fellow in 2022.
\end{IEEEbiography}
\end{document}

%% file: 1_Introduction.tex
\IEEEPARstart{R}{ecent} advancements in sensing and computation are driving a modern healthcare revolution. 
Minimally invasive surgery (MIS) has specifically proven to be a promising application domain for robotics and artificial intelligence~\cite{Chadebecq2023} since visual and physical access to the patient's organs are so restricted.
Surgical (semantic) scene segmentation and workflow estimation~\cite{Kitaguchi2022,Garrow2021} are at the forefront of this effort due to the increasing availability of laparoscopic video recordings and crowdsourced labeling methods. Novel sensing techniques such as hyper- and multi-spectral imaging are pushing the boundary of intraoperative functional evaluation and diagnosis~\cite{Clancy2020SurgicalImaging}.
Large language models and their connection with advanced robotic planning~\cite{Huang2023} are suggesting an imminent acceleration in human-machine interaction that would reshape the concepts of prediction and assistance in surgery~\cite{Bhattacharya2023}.

\begin{figure}[t!]
    \centering
    {\includegraphics[width=\columnwidth]{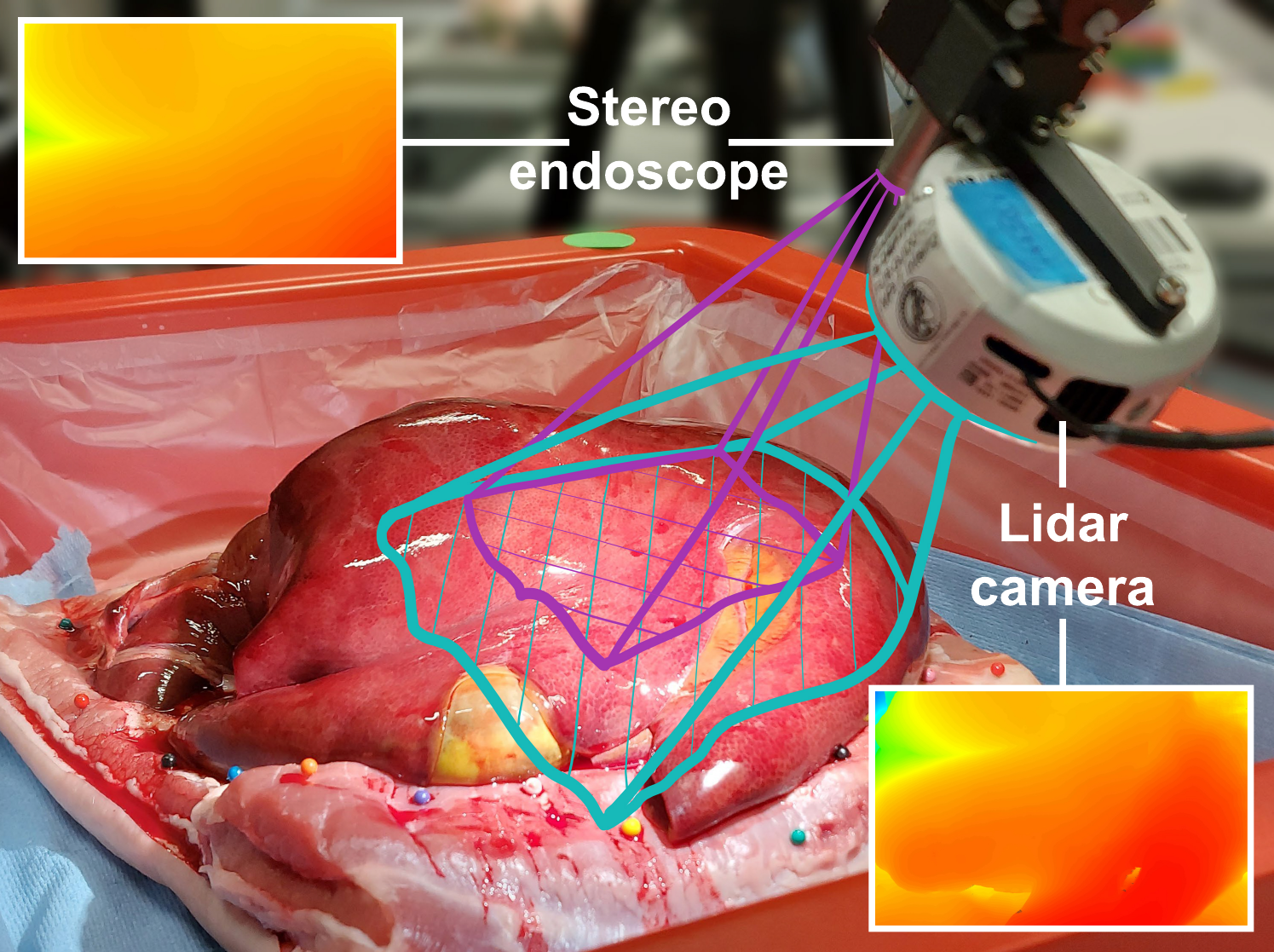}}
    \caption{Fresh ex-vivo porcine tissue being imaged by both a standard stereo endoscope and a rigidly attached compact \lidar{} (light detection and ranging) camera. The inset images show the depth maps seen by each camera. This sensor setup was used to capture all study data.}
    \label{Fig:cover}
\end{figure}

In this quickly evolving panorama, we believe \textit{geometrically exact 3D representations of the patient's anatomy} and their \textit{efficient, continuous, and reliable generation} are key elements for the successful integration of more advanced technology into surgery.
Depth estimation and 3D reconstruction in this challenging context are therefore to be seen as a foundation for the development of vision-based surgical assistance. 
Appropriately visualized, an accurate real-time map of the anatomical workspace could potentially improve the 3D scene understanding of the surgeon and also open the way for partially or fully autonomous procedures.

More broadly, the concurrent use of multiple sensing modalities, from traditional approaches based on visible light, ultrasound, and x-rays to experimental methods such as hyperspectral or photo-acoustic imaging, is currently limited by the lack of efficient data fusion methods and intuitive visualization interfaces. 
Allowing for the simultaneous display of anatomical, functional, and operational information would require real-time registration and (relative) pose estimation in 3D space. Such a registration process is possible only if a robust and accurate reconstruction of the surgical cavity is available and constantly updated.
The adoption of 3D sensing technologies for real-time intraoperative imaging remains, however, relatively limited in current clinical practice~\cite{Amparore2022}. The challenges are multi-faceted and depend on the specific surgical application. 

As reviewed in Section~\ref{sec:related_work_image}, monocular and stereo-based depth estimation in the visible light spectrum represents the state of the art for shape sensing in endoscopy, driven and accelerated by the advent of deep learning.
While achieving excellent performance in many applications, the output of each model still highly depends on the image content and the availability of training data, rendering its setup tedious and its utility limited in scope. Furthermore, for the deployment of large state-of-the-art models, processing time still remains a bottleneck.
Consequently, clinically viable methods for collecting real-time depth measurements during surgery with reference-level accuracy are still missing. 

In the last decades, laser-based time-of-flight (ToF), also known as light detection and ranging (\lidar{}), has emerged as a leading technology for indoor and outdoor robotic perception. In particular, \lidar{} sensors provide long-range, high-accuracy measurements of the 3D environment. Research and commercial applications for \lidar{} in mobile robotics span from autonomous cars~\cite{Chen2021RangeVehicles}, aerial vehicles~\cite{Yao2019UnmannedReview,khattak2020}, and construction machines~\cite{Nubert2022Graph-basedRobots} to more recent frameworks for deployment on legged robots~\cite{nubert2022learning}.
While sensor size and weight matter less in the aforementioned applications, the introduction of such technology in surgery has mostly been limited by device size, heat production, and minimum imaging distance. Yet, the recent introduction of \lidar{} in hand-held consumer devices, such as smartphones and tablets, shows a promising perspective toward high-resolution, short-range, miniaturized \lidar{} sensors that are well-suited for future use in surgical applications.  Section~\ref{sec:related_work_tof} summarizes past research in this direction.

This study rigorously investigates the use of a small commercial \lidar{} sensor for imaging different ex-vivo biological tissues at surgery-relevant working distances. Section~\ref{sec:rationale} presents the rationale for our experiment, and Sections~\ref{sec:method}, \ref{sec:processing}, \ref{sec:experiments}, and \ref{sec:results} present the materials, computational methods, procedures, and results of an extensive multifactorial performance comparison between the 3D output of the \lidar{} and a state-of-the-art learning-based stereo matching algorithm based on endoscopic images.
Our contributions are as follows:
\begin{itemize}
    \item A detailed description of an \textbf{experimental procedure} for acquiring, processing, and evaluating sensor point clouds (against the measured ground-truth shape), allowing multifactorial \textbf{quantitative analysis on signed error fields} using rigorous statistical tools.
    \item A detailed report on how near-infrared time-of-flight imaging (\lidar{}) and traditional 3D endoscopy (based on deep stereo matching) are affected by \textbf{geometric, biological, and optical factors} that are relevant for surgery.
\end{itemize}
Based on these contributions, we conclude with a critical discussion of the limitations of current approaches to 3D intraoperative imaging in minimally invasive surgery in Section~\ref{sec:discussion}, followed by future perspectives for multi-viewpoint real-time 3D reconstruction in Section~\ref{sec:conclusion}.

%% file: 2_RelatedWork.tex
\subsection{Image-based Depth Estimation}
\label{sec:related_work_image}
For decades, the most widely used technique for depth estimation from images has been geometric stereo matching, with semi-global matching (SGM) being among the most popular approaches~\cite{Hirschmuller2005AccurateInformation} for dense depth reconstruction from a rectified stereo pair. By using two calibrated cameras and by knowing their relative displacement and orientation, one can estimate a disparity value between the left and right view for every pixel in the image space. The disparity is then converted into depth using the intrinsic camera parameters and stereo baseline, as depth is inversely proportional to the disparity. For a number of applications, endoscopic imaging included, the use of stereo cameras has long been seen as impractical or unnecessary, limiting the majority of available data to a single RGB view. 
Few classical (i.e., non-learning-based) attempts have been made to estimate depth from a single image~\cite{Saxena2009Make3D:Image, Ladicky2014PullingPerspective}, as the problem is inherently ill-posed. The widespread adoption of modern (deep) learning methods, such as convolutional neural networks (CNNs)~\cite{lecun1995convolutional}, and the increasing availability of indoor and outdoor ground-truth datasets~\cite{Khan2021AEstimation} have rapidly accelerated depth estimation for both monocular~\cite{Ming2021DeepReview,Masoumian2022MonocularReview} and stereo images~\cite{Poggi2022OnSurvey,Laga2022AEstimation}. It is to be expected that novel architectures such as image transformers~\cite{Dosovitskiy2021} and diffusion models~\cite{Rombach_Diffusion_2022_CVPR}, as well as the introduction of more generic foundation models~\cite{oquab2023dinov2}, will continue advancing the future of stereo matching and depth estimation~\cite{Cheng2021Swin-Depth:Estimation,Li2021RevisitingTransformers}.\looseness-1

Stereo matching faces additional challenges in the surgical setting compared to the traditional application domains of computer vision. Endoscopic images frequently suffer from low light; a lack of distinctive visual features; the presence of blood, smoke, and reflective surfaces; limited resolution; and a small baseline between the left and right cameras~\cite{Ward2021ComputerSurgery}. To solve this last issue, \mbox{Avinash et al.} proposed an insertable pick-up stereo camera that allows for an increased stereo baseline and improved maneuverability from a second viewpoint~\cite{Avinash2019AStudy}; however, the tested camera electronics did not allow for actual use through a standard surgical cannula, and no 3D performance assessment was reported. 

Moreover, training deep-learning algorithms results in further challenges due to the scarcity of publicly available surgical datasets in general, and the frequent lack of ground-truth depth measurements in particular. Akin to other domains, the accuracy and robustness of image-based learning methods for depth estimation strictly depend on the quality of the training data used~\cite{Mascagni2022ComputerValue}. 
For the first datasets specific to endoscopic surgery, ground truth is not available, as most of the images were recorded in the operating room on living human patients~\cite{hamlyn}. In more recent phantom and ex-vivo datasets, the 3D reference surface was generated using external hardware, like a laser scanner~\cite{Penza2018EndoAbSAlgorithms}, a structured light projector coupled with a stereo endoscope~\cite{Allan2021StereoChallenge}, or an x-ray machine~\cite{Edwards2021SERV-CT:Reconstruction}. As an alternative to in-vivo experiments and wet-lab data collection, which are often complex and resource-intensive, simulation is gaining a prominent role in the generation of datasets with ground truth. Long-sequence realistic simulated datasets~\cite{Cartucho2021VisionBlender:Surgery}, frequently augmented by the use of GANs~\cite{goodfellow2020generative} and soon predictably diffusion models~\cite{Rombach_Diffusion_2022_CVPR}, are pushing forward supervised and unsupervised learning approaches~\cite{Rau2019ImplicitEndoscopy,Huang2021Self-supervisedImages} in the domain of computer-assisted surgery. 

Computational efficiency represents a further challenge for real-time perception. A recent trend in this direction is the combination of 3D reconstruction with other surgical computer-vision tasks, such as tool segmentation or tissue tracking, in synergistic (and ideally more efficient) network architectures~\cite{Psychogyios2022MSDESIS:Segmentation,Lin2022Semantic-SuPer:Tracking}. Nevertheless, the processing of high-resolution raw images to produce 3D point clouds still introduces significant delays. Even with efficient parallel GPU programming and high computational power, two-dimensional stereo imaging for depth estimation imposes a trade-off between accuracy and run-time performance. Thus, as a reference for stereo matching, we selected RAFT-Stereo~\cite{Lipson2021RAFT-Stereo:Matching}, a state-of-the-art learning-based method based on optical flow. At the time of the experiments, it represented one of the best-performing publicly available stereo algorithms capable of real-time processing~\cite{burrus}.

\subsection{Structured Light and Time-of-Flight Sensing}
\label{sec:related_work_tof}
Another sector of research investigates the use of specialized hardware for surface reconstruction.
Over the years, several research groups proposed the use of structured-light projectors to enhance shape estimation in endoscopic or laparoscopic imaging systems~\cite{Maurice2012ASurgery, Schmalz2012AnLight, Edgcumbe2014PicoSurgery, Reiter2014SurgicalImaging, Le2018DemonstrationAnastomosis}.
As a remarkable example, Maurice et al.\ presented a system capable of integrating light patterns into a full HD endoscopic stream at 25~Hz. The usability of the system was tested in vivo on a pig abdomen, but no quantitative evaluation of the 3D reconstruction was reported~\cite{Maurice2012ASurgery}. Overall, structured-light technology has shown limited robustness and applicability to the clinical workflow~\cite{Stolyarov2022Sub-millimeterFlight}.
As a more practical extension, Weld et al.\ recently proposed a learning-based solution to stereo matching where disparity and structured light patterns are estimated jointly~\cite{Weld2022}.

Penne et al.\ first presented the concept of time-of-flight 3D endoscopy~\cite{Penne2009Time-of-FlightEndoscopy}. However, the selected laser was not eye-safe, and the reconstructed volume was limited to a few thousand points. 
A subsequent work proposed a method for fusing ToF sensing with standard RGB camera images for augmented 3D endoscopy~\cite{Haase2013DEndoscopy}. The introduction of the RGB information enhanced the robustness of the 3D reconstruction and allowed for the localization of surgical tools. More recently, Roberti et al.\ presented a prototype of an endoscope with a chip-on-tip ToF design to overcome the narrow view field introduced by the fiber optic coupling~\cite{Roberti2021AReconstruction}; its limitations are mostly related to size, resolution, and thermal dissipation. 

Stolyarov et al.\ customized an off-the-shelf \lidar{} to be coupled to a commercial endoscope~\cite{Stolyarov2022Sub-millimeterFlight}. To the best of our knowledge, this solution represents the most advanced research implementation of \lidar{} sensing for endoscopy. Still, the resolution is limited to $640\times480$ pixels, the proposed hardware setup is relatively cumbersome to build and use, and several steps of post-processing are necessary, such as spatial and temporal filtering as well as compensation for fiber optic distortion. Quantitative comparison with a laser scanner on ex-vivo porcine kidneys showed nearest neighbor average errors in the \SI{0.75}{\mm} to \SI{10}{\mm} range~\cite{Stolyarov2022Sub-millimeterFlight}. 
Interestingly, Stolyarov et al.\ reported the need for testing ToF technology with alternative hardware solutions, at various working distances, and with different tissue targets, speculating that \lidar{} measurement accuracy might depend on tissue type~\cite{Stolyarov2022Sub-millimeterFlight}.

%% file: 3_Rationale.tex
Our experimental setup and evaluation scheme were carefully designed to enable a thorough first-of-its-kind comparison between \lidar{} imaging and stereo endoscopy on biological tissue. In contrast to our earlier work~\cite{Caccianiga2022DenseSurgery}, this study is performed on real animal tissue, adds more investigated scenarios, and constitutes a far more thorough (statistical) analysis and evaluation. Our goal is to \textit{characterize the performance of two different real-time intraoperative imaging techniques while varying selected surgery-relevant experimental conditions}. Importantly, we sought to assess the produced point clouds in a systematic, reproducible, and quantitative way.

Section~\ref{sec:method} describes the experimental setup, including the selected ex-vivo tissue samples, our two cameras, and the tool we used to capture an accurate ground-truth 3D representation of each tissue sample. It also describes the pipeline we created to generate, acquire, and pre-process the data. Section~\ref{sec:processing} presents the method used to link the kinematic space of the cameras (the 3D output) and the subject (the ground-truth shape), a set of performance metrics derived from the comparison between the 3D output and the ground truth, and a set of methods to evaluate and visualize our performance metrics, both qualitatively and quantitatively. Though contextualized in our specific setup, these methods have broader research relevance since they aim to solve \textit{challenges that are common to robotics and 3D vision}, such as camera-to-world calibration and quantitative evaluation of point clouds; therefore, they can easily be generalized to other application domains.

%% file: 4.1_Materials.tex
\begin{figure*}[t]
    \centering
    \includegraphics[width=\textwidth]{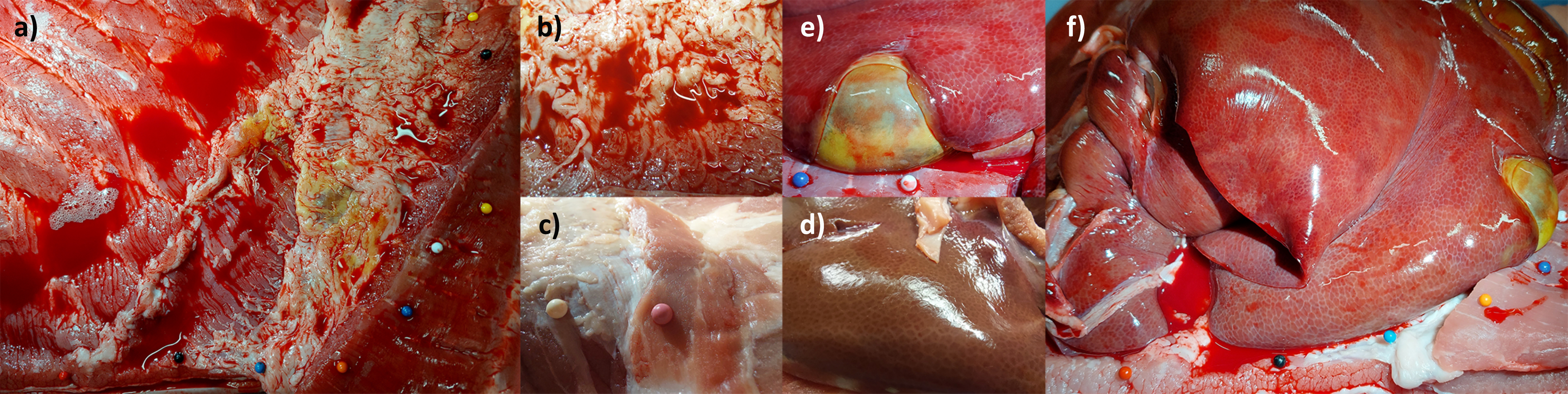}
    \vspace{-.7cm}
    \caption{Ex-vivo porcine tissue samples. a) Abdomen: overall, bloody. b) Abdomen: fatty tissue, bloody. c) Abdomen:  muscular tissue, clean. d) Liver: smooth surface, clean. e) Liver: gallbladder, bloody. f) Liver: overall, bloody.}
    \label{Fig:organs}
\end{figure*}

\begin{figure*}[t]
    \centering
    \vspace{-.2cm} 
    \includegraphics[width=\textwidth]{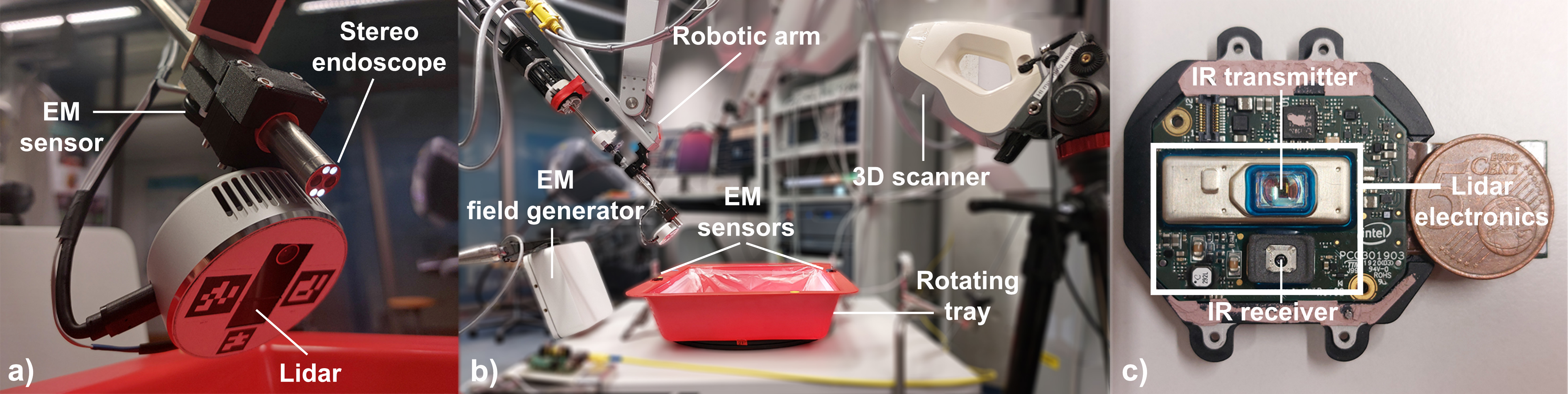}
    \vspace{-.7cm}
    \caption{Experimental setup. a) Main components rigidly attached to the da Vinci robot's camera arm: electromagnetic (EM) sensor, stereo endoscope, and \lidar{}. b) Overall hardware setup. c) \Lidar{} electronics exposed to highlight the infrared (IR) transmitter and receiver, with a five-euro-cent coin for size reference.}
    \label{Fig:setup}
\end{figure*}

\subsection{Ex-vivo Tissue}

Two fresh samples of ex-vivo animal tissue were used for the experiments: \textit{i)} a square cut from the porcine abdominal wall ($30$\,cm\,$\times$\,$30$\,cm\,$\times$\,$5$\,cm, Fig.~\ref{Fig:organs}a--c), and \textit{ii)} a whole porcine liver with intact gallbladder and bile ducts (approximately $25$\,cm\,$\times$\,$20$\,cm\,$\times$\,$15$\,cm, Fig.~\ref{Fig:organs}d--f). The two samples were preserved at a controlled temperature in sealed plastic bags with their physiologically secreted fluids until the experiments to preserve their moisture level and surface characteristics. To simulate a surgical field with bleeding, a sample of fresh porcine blood ($500$\,ml) was also obtained to be poured on the tissue for particular experiment conditions.

Porcine models were chosen to achieve high similarity with human anatomy. The abdominal wall sample resembles the combination of fat (Fig.~\ref{Fig:organs}b) and muscular tissue (Fig.~\ref{Fig:organs}c) that can be seen in the background of minimally invasive laparoscopic procedures. The liver frequently appears in MIS scenes due to its size and involvement in high-volume procedures such as cholecystectomy, pancreatectomy, and liver lobectomy. The liver is also an interesting subject from the computer-vision perspective due to its particularly smooth and monochromatic surface (Fig.~\ref{Fig:organs}d). 

\subsection{Experimental Apparatus}
To perform the current experiments, we used a da Vinci Si Surgical System (Intuitive Surgical Inc., Sunnyvale, USA) equipped with a $0^\circ$ HD stereo endoscope (Fig.~\ref{Fig:setup}a) with an adjustable light source. Furthermore, we deployed a RealSense L515 \lidar{} camera (Intel Corporation, Santa Clara, USA) rigidly attached near the end of the endoscope shaft (Fig. \ref{Fig:setup}a). In this way, the two cameras have the same distance and perspective when observing an object, so their 3D reconstruction accuracy can be fairly compared. 

We placed the abdominal wall sample of animal tissue into a rectangular plastic tray (Fig.~\ref{Fig:setup}b) to be imaged. An Artec Eva 3D scanner (Fig.~\ref{Fig:setup}b) that uses white structured-light technology was used to obtain an accurate ground-truth reconstruction of this scene. The liver was later laid on top of the abdominal wall and reconstructed with the scanner as well. A circular base with bearings allows for smooth $360^\circ$ rotation around the vertical axis. This rotation mechanism allows for precise re-orientation of the tray with respect to the cameras while preserving the shape of the tissue after the ground-truth 3D scan was captured. To keep track of the position and orientation of the rotating tray relative to the cameras, we used an Aurora electromagnetic tracking system (Northern Digital Inc., Ontario, Canada): electromagnetic sensors were attached to the tray (Fig.~\ref{Fig:setup}b) and the da Vinci camera arm (Fig.~\ref{Fig:setup}a). A $36$-core Intel i9 desktop PC running Ubuntu 20.04 with an Nvidia RTX 3080 GPU was used for hardware control, signal processing, and data recording. 

\subsection{Data Acquisition and Pre-processing}
\label{Sect:acquisition}
\subsubsection{Stereo Matching} The left and right image pairs produced by the stereo endoscope (FHD: $1920\times1080$, 30~Hz) are intercepted by a DeckLink Quad 2 acquisition card (Blackmagic Design, Port Melbourne, Australia). Custom DeckLink drivers~\cite{Forte2021DesignLymphadenectomy} provide the endoscope frames as raw image topics on the local Robot Operating System (ROS) network. The preprocessing steps of rectification and spatial down-sampling (reduction to nHD: $640\times360$) are performed on the raw images. Stereo matching is implemented with a custom ROS wrapper around the fastest CUDA version of RAFT-Stereo~\cite{Lipson2021RAFT-Stereo:Matching}. This algorithm outputs disparity maps at approximately 5~Hz when operating on the FHD frames, and it reaches about $15$~Hz on the down-sampled images. The nHD resolution is chosen for the current experiments to grant as much real-time performance as possible. The disparity maps are converted to depth maps and then re-projected as 3D point clouds.\looseness-1

\subsubsection{\Lidar{} Stream} 
The raw measurements produced by the L515 \lidar{} are acquired on Ubuntu through the RealSense SDK (v2.50) and the RealSense ROS wrappers (v2.3.2, on ROS Noetic)\footnote{\url{ https://github.com/IntelRealSense/realsense-ros/tree/2.3.2}}. The L515 has an out-of-the-box minimum imaging distance of $0.25$\,m. Through the SDK, we program a customized configuration to allow for time-of-flight depth imaging at close range ($\ge0.05$\,m). The point-cloud stream is directly available on the ROS network as raw 3D output ($1024\times768$, $30$~Hz). The L515's embedded RGB camera also produces full HD monocular images ($1920\times1080$, $30$~Hz). Both the \lidar{} depth and RGB streams are down-sampled and cropped to $640\times360$ to match the RAFT-Stereo output.

\subsubsection{Ground Truth} 
We acquire a $360^\circ$ scan of each sample of animal tissue in the tray with the Artec 3D scanner before each experiment configuration. Artec Studio 16\footnote{\url{https://www.artec3d.com/3d-software/artec-studio software}} is used for both acquisition and post-processing of the ground-truth (GT) scans. Partial scans of the same object are registered and fused, cleaned up, and color-textured. The resulting meshes are then exported in .ply format and imported via a Python script for comparison with the sensed 3D data streams. We performed a sanity check on the GT scans by measuring the tissue tray's major dimensions with a precision caliper. Comparing with the same dimensions in sample scans showed no noticeable deformation, matching Artec's 3D accuracy specifications ($0.1$\,mm + $0.3$\,mm/m).

\subsubsection{Electromagnetic Tracking} 
The Aurora tracking system is connected to the PC via USB, and the six-degree-of-freedom (6-DoF) poses of the electromagnetic sensors are captured through the \texttt{sawNDITracker}\footnote{\url{https://github.com/jhu-saw/sawNDITracker}} ROS package at $50$\,Hz.

%% file: 4.2_Processing.tex
\begin{figure}[t]
    \centering
    \includegraphics[width=\columnwidth]{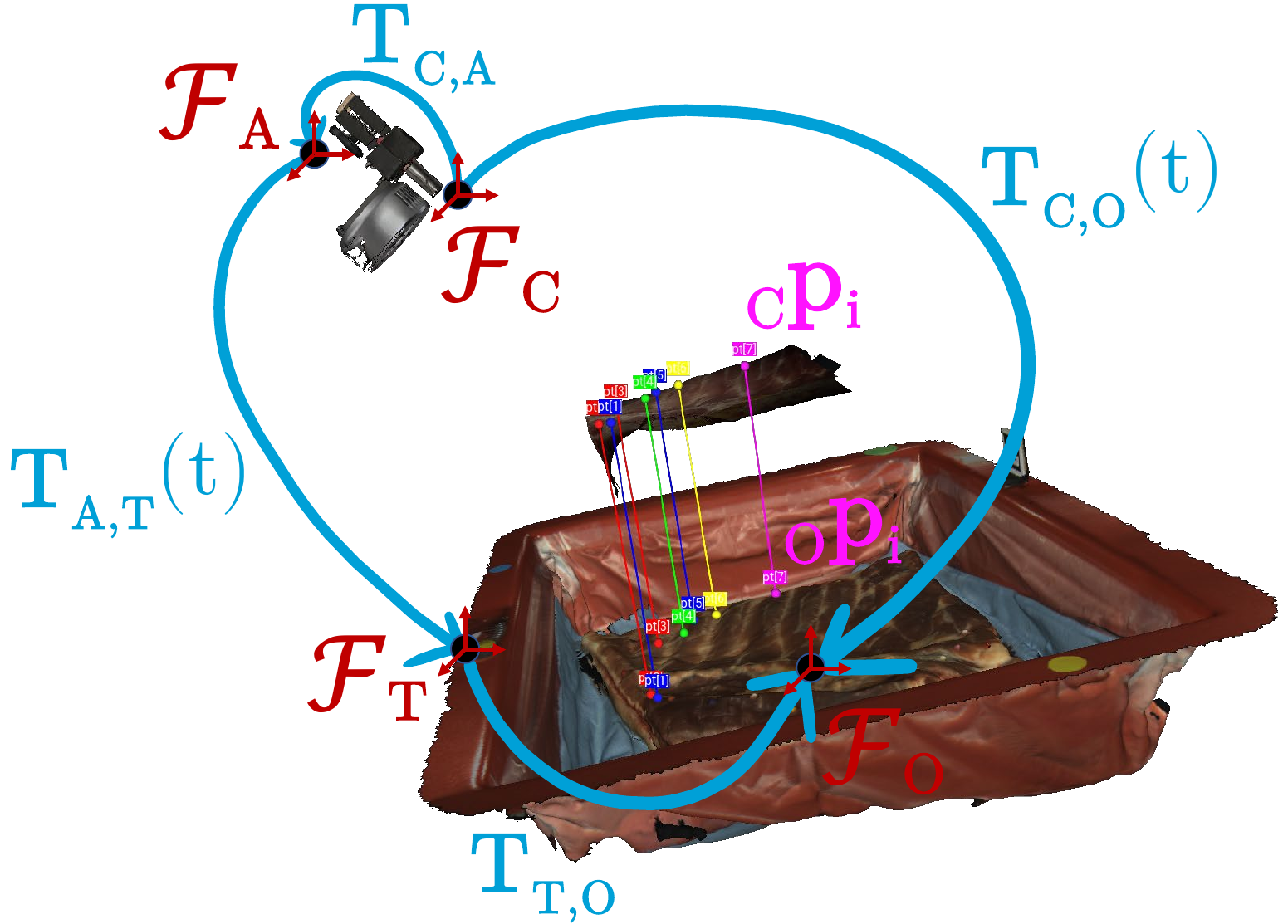}
    \caption{Kinematic transformations (arrows in light blue) between the camera ($\mathcal{F}_{\mathtt{C}}$), the robotic arm ($\mathcal{F}_{\mathtt{A}}$), the rotating tray ($\mathcal{F}_{\mathtt{T}}$), and the GT scan ($\mathcal{F}_{\mathtt{O}}$), with each frame illustrated in red. The center of the illustration also shows a sample manual pairing of all the visible plastic pins selected on the GT scan ($_{\mathtt{O}}\mathbf{p}_i$) and on a corresponding measured point cloud ($_{\mathtt{C}}\mathbf{p}_i$).}
    \label{Fig:kinematics}
\end{figure}

\subsection{Endoscope Stereo Calibration} 
\label{calibrations}
The intrinsic parameters of the left and right cameras of the endoscope are necessary to perform image rectification and to transform depth maps into point clouds. Furthermore, the extrinsic transformation between the two cameras of the endoscope is needed to transform the disparity maps into depth maps. All of these transformation matrices are obtained through the ROS native stereo calibration package\footnote{\url{http://wiki.ros.org/camera_calibration/Tutorials/StereoCalibration}}. A rectangular checkerboard ($5\times7$ squares, each $10$\,mm\,$\times$\,$10$\,mm) is moved across the endoscope's field-of-view at an axial distance of $5-20$\,cm. An epipolar re-projection error of $0.18$ pixels was obtained for our chosen calibration output.

\subsection{Camera-to-world Registration} 
To produce accuracy and precision metrics for the image sensors, one must know the kinematic transformation between each produced point cloud and its respective ground-truth scan. This transformation needs to be re-calculated every time the camera or the observed tissue moves. 
We approach this registration step as a \textit{camera-to-world} pose-estimation problem. 
As visualized in Fig.~\ref{Fig:kinematics}, we define the four coordinate frames $\mathcal{F}_{\mathtt{O}}$, $\mathcal{F}_{\mathtt{T}}$, $\mathcal{F}_{\mathtt{A}}$, and $\mathcal{F}_{\mathtt{C}}$, for the \textbf{organ} (origin of the ground-truth 3D scan of the imaged tissue), \textbf{tray} (EM sensor), \textbf{arm} (EM sensor), and \textbf{camera}, respectively. 
Here, $\mathbf{T}_{\mathtt{T},\mathtt{O}}$ and $\mathbf{T}_{\mathtt{C},\mathtt{A}}$ denote the static but initially unknown transformations, while $\mathbf{T}_{\mathtt{A},\mathtt{T}}(t)$ denotes the dynamic transformation measured by the electromagnetic tracker. 
The kinematic relationship is
\begin{equation}
    \mathbf{T}_{\mathtt{C},\mathtt{O}}(t) = \mathbf{T}_{\mathtt{C},\mathtt{A}}\cdot \mathbf{T}_{\mathtt{A},\mathtt{T}}(t)\cdot\mathbf{T}_{\mathtt{T},\mathtt{O}}
\label{camera2world}
\end{equation}
with $\mathbf{T}_{\mathtt{\bullet},\mathtt{\bullet}} \in SE(3)$ for all transformations.

In a traditional robotic approach, $\mathbf{T}_{\mathtt{C},\mathtt{A}}$ is calculated via hand-to-eye calibration, $\mathbf{T}_{\mathtt{A},\mathtt{T}}(t)$ is equivalent to the forward kinematics (FK) calculated with respect to the robot base, and $\mathbf{T}_{\mathtt{T},\mathtt{O}}$ is usually the unknown base-to-world transformation. In robotic surgery settings, the camera-to-world kinematic loop is often closed by bringing an end-effector to specific reference landmarks that are also visible in the calibrated camera view field~\cite{Kalia2019Marker-lessReality}; such an approach is subject to the availability and accuracy of the FK, which we do not have.

For our experiments, we identified and implemented two distinct ways to obtain the camera-to-world registration, depending on the experimental constraints. \textit{i)} The first method ($\mathbf{T}_{\mathtt{C},\mathtt{O}}^{\mathrm{pins}}$) is based on spherical fiducial pins placed on the surface of the tissue. The pins act as visible anchor points in a systematic and reproducible manner, independent of the tissue morphology or any interaction between the imaging technology and the biological tissue. \textit{ii)} The second method ($\mathbf{T}_{\mathtt{C},\mathtt{O}}^{\mathrm{kine}}$) combines the calibration obtained from the pins with measurements from the EM tracker when less than four fiducial pins are visible in the scene (e.g., in a close zoom setting).
The following subsections detail these two approaches.

\subsubsection{Fiducial-based Registration ($\mathbf{T}_{\mathtt{C},\mathtt{O}}^{\mathrm{pins}}$)}
24 colored spherical pins ($\diameter_{\mathrm{head}} = 4$\,mm) were inserted into the upper surface of the abdominal wall.  
For each captured frame, we manually pair pins (Fig.~\ref{Fig:kinematics}) visible in the ground-truth scan, $_{\mathtt{O}}\mathbf{p}_i$, with all pins visible in the selected camera's point cloud $_{\mathtt{C}}\mathbf{p}_i$, for $i \in \{1,\dots,24\}$. To reduce the influence of high-frequency noise, we perform this pin pairing on camera point clouds that are averaged across 125 successive static frames.
Following the standard point-cloud registration convention used in~\cite{tuna2022x}, we refer to the resulting point matching as $\mathcal{M} \in \mathbb{R}^{6 \times N} = \text{matching}\big(\, _{\mathtt{O}}\boldsymbol{P},\ _{\mathtt{C}}\boldsymbol{P}\big)$, with $N$ denoting the number of found matches.
Given this matching for at least four pins, we can obtain the transformation $\mathbf{T}_{\mathtt{C},\mathtt{O}}^{\mathrm{pins}}(t)$ using singular value decomposition (SVD) for a specific time $t$ of the camera point-cloud recording. The manual matching and computation of the transformation are both performed using Cloud Compare\footnote{\url{http://www.cloudcompare.org/}}.

\begin{figure*}[t!]
    \centering
    \includegraphics[width=\textwidth]{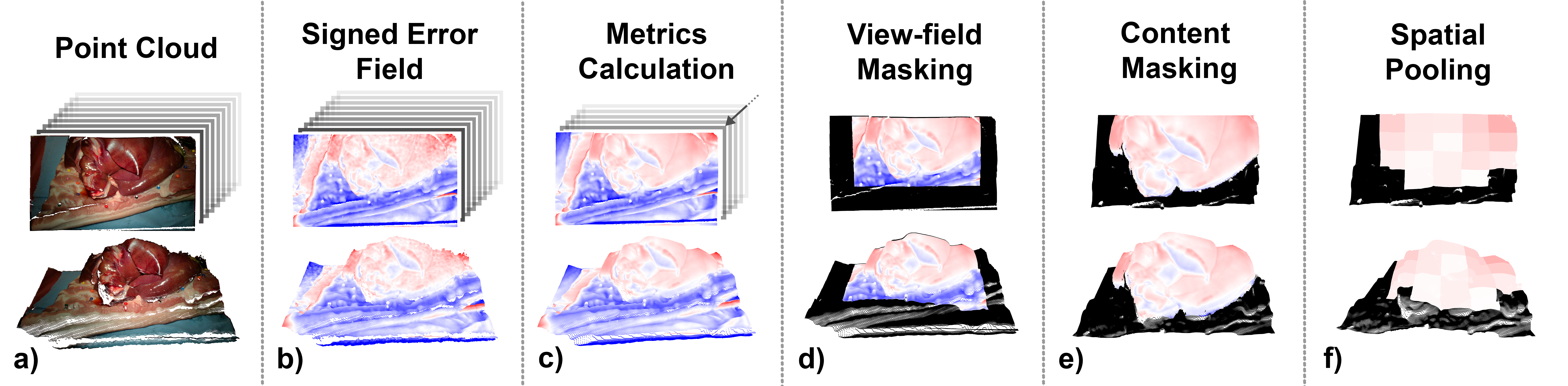}
    \vspace{-.7cm}
    \caption{Point-cloud post-processing steps (Section~\ref{metrics}) applied to a sample \lidar{} frame. a) Raw colored point cloud. b) Signed error field as distance from ground truth. c) Error averaged across multiple static frames. d) Mask to equalize the view field across cameras. e) Mask to extract tissue-specific points. f) Error values pooled (averaged) across space to reduce the resolution for statistical analysis.
    Point clouds from the stereo endoscope undergo the same process.}
    \label{Fig:tiling}
\end{figure*}

\subsubsection{Kinematics-based Registration ($\mathbf{T}_{\mathtt{C},\mathtt{O}}^{\mathrm{kine}}$)}
\label{sec:kine}
whenever too few pins ($N < 4$) are visible in the acquired frame, the transformation $\mathbf{T}_{\mathtt{C},\mathtt{O}}$ needs to be computed from the terms on the right side of Eq.~\eqref{camera2world}. For our experiments, the origin of the ground-truth scan $\mathcal{F}_{\mathtt{O}}$ in Artec Studio is set to be coincident with the origin of the electromagnetic sensor $\mathcal{F}_{\mathtt{T}}$ attached to the tray. Thus, $\mathbf{T}_{\mathtt{T},\mathtt{O}}$ is the identity matrix ($\mathbf{I}$). From the closest camera pose ($t=t_p$) where calibration was performed via fiducial pins, we exploit Eq.~\eqref{camera2world} with $\mathbf{T}_{\mathtt{C},\mathtt{O}}^{\mathrm{pins}}(t_p)$ and the corresponding EM measurement $\mathbf{T}_{\mathtt{A},\mathtt{T}}(t_p)$ to compute $\mathbf{T}_{\mathtt{C},\mathtt{A}}(t_p)$, as follows:
\begin{equation}
    \mathbf{T}_{\mathtt{C},\mathtt{A}}(t_p) = \mathbf{T}_{\mathtt{C},\mathtt{O}}^{\mathrm{pins}}(t_p)\cdot\mathbf{I}\cdot\mathbf{T}_{\mathtt{T},\mathtt{A}}(t_p).
\label{local_hand-eye}
\end{equation}
By closing the kinematic loop in this way, we make sure that any error linked to the manual alignment of $\mathcal{F}_{\mathtt{O}}$ and $\mathcal{F}_{\mathtt{T}}$ is implicitly considered in $\mathbf{T}_{\mathtt{C},\mathtt{A}}$, which is locally exact at the instant $t_p$, up to the accuracy of $\mathbf{T}_{\mathtt{C},\mathtt{O}}^{\mathrm{pins}}(t_p)$. 

As a result, when the camera is in a configuration for which the pins are not visible, $\mathbf{T}_{\mathtt{C},\mathtt{O}}^{\mathrm{kine}}(t_{k})$ is computed by simply substituting $\mathbf{T}_{\mathtt{C},\mathtt{A}}(t_p)$ into Eq.~\eqref{camera2world}:

\begin{equation}
    \mathbf{T}_{\mathtt{C},\mathtt{O}}^{\mathrm{kine}}(t_k) = \mathbf{T}_{\mathtt{C},\mathtt{A}}(t_p)\cdot \mathbf{T}_{\mathtt{A},\mathtt{T}}(t_k)\cdot\mathbf{I}.
\label{camera2world_2}
\end{equation}
Therefore, $\mathbf{T}_{\mathtt{C},\mathtt{O}}^{\mathrm{kine}}(t_k)$ relies on the accuracy of $\mathbf{T}_{\mathtt{C},\mathtt{O}}^{\mathrm{pins}}(t_p)$ as well as that of the EM tracker,  which depends on the magnitude of the motion that occurred between $t_p$ and $t_k$.

\subsection{Offline Post-processing}
\label{metrics}
Figure~\ref{Fig:tiling} illustrates the five post-processing steps performed on raw point-cloud data gathered from the cameras.

\subsubsection{Signed Error Field}
\label{error}
To quantitatively evaluate the 3D output of each camera, we compute the signed error (distance) between each point in the point cloud and the respective GT mesh measured with the 3D scanner. 
A Python script processes each frame of the recorded ROS bag files, first performing the camera-to-world (tissue) registration using $\mathbf{T}_{\mathtt{C},\mathtt{O}}^{\mathrm{pins}}$ or $\mathbf{T}_{\mathtt{C},\mathtt{O}}^{\mathrm{kine}}$. For each measured point, it searches for the three closest vertices on the GT mesh. It then computes the normal of the triangular face created by the three vertices, calculates the point's distance from the center of the face, and projects the distance along the normal to obtain the sign; positive means it is closer to the camera than the GT. As a result, for each point in the cloud (Fig.~\ref{Fig:tiling}a), we extract the signed distance from the GT (Fig.~\ref{Fig:tiling}b) and the global coordinates of the corresponding triangle center. 

\subsubsection{Metric Calculation}
While processing the bag files, we extracted $n = 125$ RGB-D (colored point cloud) frames for each experimental condition, corresponding to $n$ successive time instances at the selected camera's acquisition frequency. 
To obtain our \textbf{Depth Accuracy} metric, we computed the per pixel \textit{mean} of the signed error across the 125 frames (Fig.~\ref{Fig:tiling}c), resulting in $640\times360$ signed scalar values from which temporal sensor noise has been removed. This metric represents the ability of the tested 3D cameras to correctly estimate the depth of each point, and its sign shows the direction of any systematic depth measurement shift. We compute our \textbf{Time Variability} metric as the \textit{standard deviation} (SD) of the signed error across the 125 frames. To obtain our \textbf{Shape Precision} metric, we begin with the Depth Accuracy metric, subtract its mean across pixels, and take the absolute value to obtain an unsigned index of local deviation from the ground-truth surface, independent of the absolute depth values measured. Fig.~\ref{Fig:averagestd} shows examples of point clouds color-mapped with the scalar fields corresponding to these three metrics. Such information is available for each camera, each tissue sample, and each experimental condition.\looseness-1

\subsubsection{View-field Masking} 
The L515 \lidar{} has a field of view that is roughly twice as large as the da Vinci stereo endoscope at an equal imaging distance. As a consequence, a portion of the \lidar{} data is not seen by the endoscope. Furthermore, especially when very close to the subject, the view field of the endoscope might include an upper portion of the tissue not seen by the \lidar{}. To compensate for these discrepancies, for each measured 3D point, we use the coordinates of the closest point on the GT (Section~\ref{error}) to find the portion of the GT surface that is seen by both cameras. The resulting mask limits further processing to this area (Fig.~\ref{Fig:tiling}d).

\subsubsection{Content Masking}
\label{sec:content_masking}
To make our signed error fields representative of only one specific tissue type, we manually segmented the GT scans to distinguish between desired (inlier) and undesired (outlier) content in the subsequent steps. This step filters out any portion of the plastic tray visible in the background and removes the surrounding abdominal tissue when evaluating reconstructions of the liver (Fig.~\ref{Fig:tiling}e).

\subsubsection{Spatial Pooling}
\label{sec:spatial_pooling}
Even though high-resolution visualization of the signed error fields is useful for qualitative assessment, we perform spatial average pooling as the last stage of post-processing to obtain a greatly reduced number of relatively independent measurements suitable for statistical analysis. Each $640\times360$ frame of error measurements is divided into 30 ($6\times 5$) rectangular tiles, and the average value in each tile is calculated (Fig.~\ref{Fig:tiling}f). If a tile is composed of more than $95$\% outlier content (e.g., the plastic tray or undesired tissue, depicted in black in Fig.~\ref{Fig:tiling}e), no value is assigned to the tile. 
As a result, for each experimental condition, the data granularity is compressed by four orders of magnitude (from $640\times360$ = $230\,400$ pixels to $30$ pooled values), allowing the resulting dataset to be processed by our statistical models without inflating the significance of the estimated p-values~\cite{Lin2013ResearchProblem}.

\subsection{Statistical Analysis} 
Our study was designed to be analyzed with a \textit{three-way repeated measures ANOVA} (analysis of variance). 
The three fixed effects are i) \textit{camera type} (Endoscope, \Lidar{}), ii) \textit{tissue type} (Abdomen, Liver), and iii) \textit{zoom} (Far, Close) for the first experiment, and i) \textit{camera type}, ii) \textit{presence of blood} (Dry, Wet), and iii) \textit{illumination} (Full, Low) for the second experiment. As described above, we extract 30 values as the output of the spatial pooling across the error fields. 
Each tile (or sample) is treated as independent, as they represent spatially separated portions of the produced RGB-D image. The tile number (1--30) is modeled as the random effect across the repeated measures. 
While exploring the data, we discovered the values were not normally distributed for any of the three error metrics. Consequently, we applied the Aligned Rank Transform (ART) by Wobbrock et al.~\cite{Wobbrock2011TheProcedures} to perform non-parametric three-way repeated-measures ANOVA and the respective post-hoc contrasts~\cite{Elkin2021AnTests}. The model was built as a generalized linear mixed model using the \textit{lmer} syntax and the \texttt{ARTool} package for R\footnote{\url{https://depts.washington.edu/acelab/proj/art/}}. ART ANOVA was performed independently for each experiment and error metric using R 4.2.2.\looseness-1

%% file: 5_Experiments.tex
We benchmark the performance of the RealSense L515 and the da Vinci Si stereo endoscope in a side-by-side comparison. First, we compare the processing delay of the two video pipelines. Then, we investigate how two pairs of experimental conditions (tissue type \& zoom, illumination \& presence of blood) affect real-time 3D reconstruction for each camera type. 

\subsection{Experiment 1: Image Processing Time}
To assess the video processing time of the \lidar{} and the da Vinci endoscope, we pointed both cameras at a monitor displaying a ROS timer with digits down to nanoseconds. The same screen also displayed the 3D video output of both cameras. 
By subtracting the time shown in each camera view (delayed) from the actual system time shown on the timer, we could assess the delay between an event happening in the surgical scene and the instant of its display to a user. The L515 generates its colored point cloud onboard; we therefore directly measured the time for the 3D image to appear on the screen. For the endoscope, we separately evaluated the three following steps: i) the native da Vinci video processing delay (time to show the event on the da Vinci external monitor), ii) the stereo pair published in ROS using the DeckLink, and iii) the final output including the stereo processing (rectification, resizing, stereo matching, and 3D reprojection). Each of these delays was measured at 10 random time points.

\subsection{Experiment 2: Effects of Tissue Type and Zoom}
To evaluate the effect of zoom, we locked all the rotational joints of the robotic arm holding the cameras while they were looking at the center of the rotating tray. Manipulation of the arm's prismatic joint allowed a pure zoom action without changing the viewing angle of the cameras. By analyzing images from a non-reported in-vivo porcine experiment, we estimated the working distances to fall in the range of $8-16$\,cm; 
similar camera distances are reported by Kalia et al.~\cite{Kalia2019Marker-lessReality} for radical prostatectomy in humans. We therefore define $16$\,cm as Far and $8$\,cm as Close. At the Far zoom setting, a sufficient number of colored pins was always visible to compute $\mathbf{T}_{\mathtt{C},\mathtt{O}}^{\mathrm{pins}}$. In contrast, enough pins were rarely visible in the Close configuration, so $\mathbf{T}_{\mathtt{C},\mathtt{O}}^{\mathrm{kine}}$ was used (Section~\ref{sec:kine}). 

For each zoom setting, we acquired eight different views of the tissue by rotating the tray $360^\circ$ at $45^\circ$ intervals. Rotating the tray highlighted a fluctuation in the accuracy of $\mathbf{T}_{\mathtt{A},\mathtt{T}}$ due to the varying distances between each EM sensor and the EM field generator (Fig.~\ref{Fig:setup}b). Computation of $\mathbf{T}_{\mathtt{C},\mathtt{O}}^{\mathrm{pins}}$ at each rotation angle (at Far) allowed us to cancel any angular error of the EM tracking when calculating $\mathbf{T}_{\mathtt{C},\mathtt{O}}^{\mathrm{kine}}$ (at Close). As a result, we minimize the drift of our camera-to-world estimation at Close, up to the accuracy of the EM tracker in measuring a linear translation of $8$\,cm. Our results are thus more robust with respect to the calibration process and do not depend on the specific part of the tissue that is visible. Our statistical analysis for this experiment treats tissue type as an independent variable. Camera viewing angle is handled not as an independent variable, but as a random effect nested in the tile variable (e.g., the same top-left tile looking at the tissue from different angles).

\subsection{Experiment 3: Effects of Illumination and Blood} 
To evaluate the effect of direct scene illumination, we dimmed the da Vinci endoscope illuminator from $100$\% to $20$\%.
The effects of low lighting, especially for endoscopes, are usually more visible at the periphery of the view field due to optical distortion and vignetting. For this reason, we did not apply the content-masking step described in Section~\ref{sec:content_masking} (Fig.~\ref{Fig:tiling}e) to measurements used for evaluating illumination.
Differentiating between types of tissue was not in the scope of this part of the experiment; consequently, this analysis treats tissue type as a random effect nested in the tile variable.

Furthermore, when considering the intraoperative presence of blood as a potential cause of error in 3D reconstruction, we hypothesized that any effects would be linked to local reflections or density changes in areas of blood accumulation, creating small pools. These areas cover only a small portion of the overall tissue surface. To remove all the other potential sources of error variation, and to maximize our chances of detecting even tiny local shape variations due to the presence of blood, we fixed the zoom at Far and kept the rotating tray steady at $0^\circ$. The registration method used throughout this experiment is thus $\mathbf{T}_{\mathtt{C},\mathtt{O}}^{\mathrm{pins}}$.

%% file: 7_Results.tex
 \begin{figure}[!t]
    \centering 
    \includegraphics[width=\columnwidth]{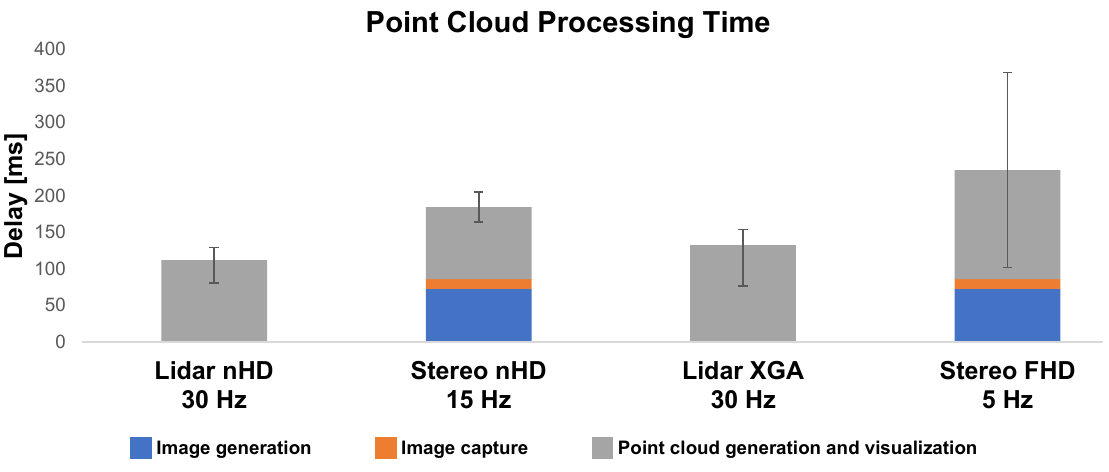}
    \caption{Results of experiment 1: mean and SD of the time delay required to produce 3D point clouds from the instant of imaging to visualization on screen for our two cameras (\lidar{} and stereo endoscope) at two resolutions.}
    \label{Fig:delay}
\end{figure}

 \begin{figure*}[tbp]
    \centering
    \includegraphics[width=\textwidth]{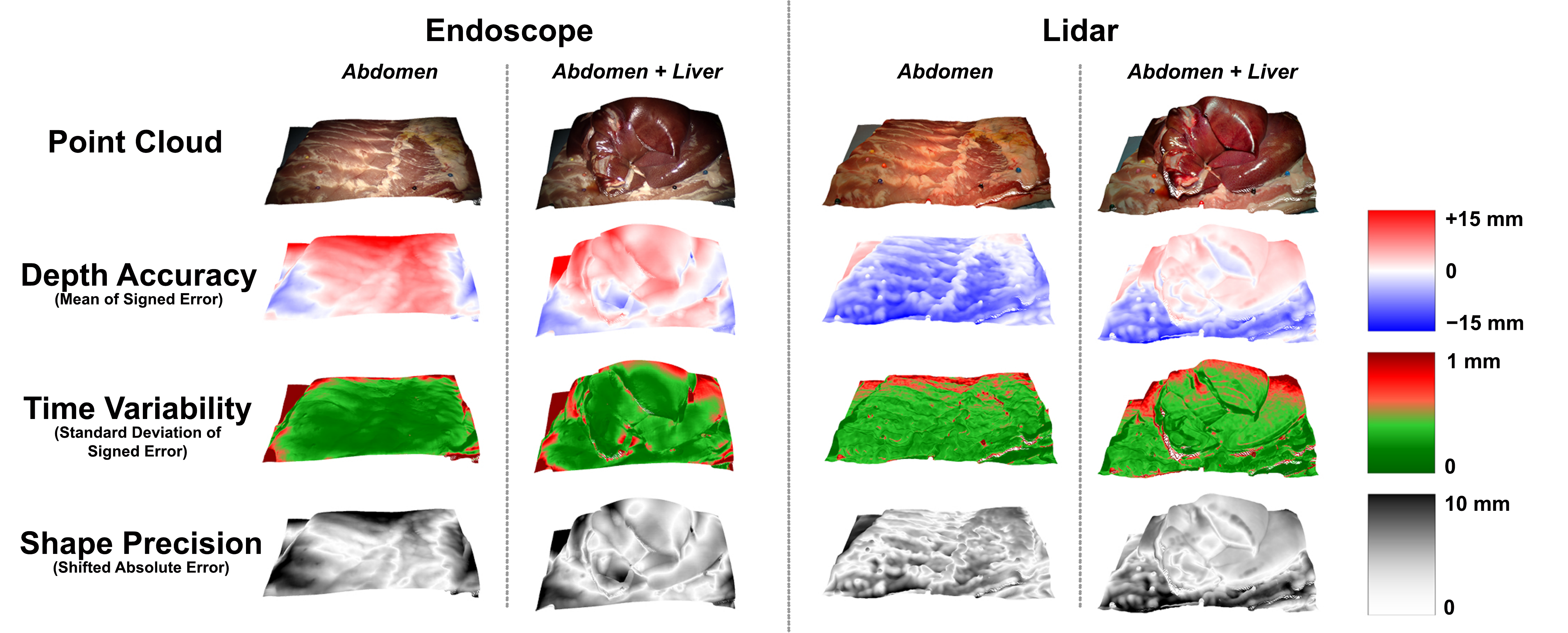}
    \vspace{-.7cm}
    \caption{Error fields computed between each measured point cloud and the respective ground-truth surface. Color mappings represent the three error metrics: Depth Accuracy (mean of signed error over time), Time Variability (SD of signed error over time), and Shape Precision (absolute error after depth shift).}
    \label{Fig:averagestd}
    \vspace{-.5cm}
\end{figure*}

\subsection{Experiment 1: Image Processing Time}
\label{time}
We compared the time to process the colored point cloud for the \lidar{} and the da Vinci stereo endoscope + RAFT both at the experiment resolution (nHD) and at the respective maximum camera resolutions (XGA for the \lidar{} and FHD for the endoscope). Fig.~\ref{Fig:delay} reports the results as mean and SD across the $10$ measurements. The \lidar{}'s total delay from measurement (in-hardware 3D imaging) to visualization (point cloud, RViz) is $112.2\pm17$\,ms at nHD and $133.1\pm20.8$\,ms at XGA. For the native da Vinci video pipeline, we measured an image generation delay of $73.3\pm26.8$\,ms. The additional time to acquire the stereo images through the DeckLink and make them available on ROS is $13\pm1$\,ms. Finally, the stereo frames are delayed due to the conversion process from raw images to point clouds (down-sampling, rectification, RAFT stereo matching, and 3D reprojection); the endoscope's resulting total delay from measurement to visualization is $185\pm20$\,ms at nHD and $235.5\pm133.2$\,ms at FHD.

\begin{table}[tbp]
\renewcommand{\arraystretch}{1.3}
\centering
\begin{threeparttable}
\caption{Experiment 2 performance comparison via ART ANOVA to investigate the effects of camera type (C), tissue type (T), zoom (Z), and their interactions.}
\label{tab:Models}
\begin{tabular}{@{}ll|ll|ll|ll@{}}
 \multicolumn{2}{c}{}  &  \multicolumn{2}{c}{\textbf{Depth Accuracy}} & \multicolumn{2}{c}{\textbf{Time Variability}} & \multicolumn{2}{c}{\textbf{Shape Precision}} \\
 \multicolumn{2}{c}{}  &  \multicolumn{2}{c}{Mean} & \multicolumn{2}{c}{SD} & \multicolumn{2}{c}{Shifted AE} \\
\toprule
Model & Df & F & $p$-value & F & $p$-value & F & $p$-value \\ \midrule
C & 1 & 99 & $<$0.001 & 1112 & $<$0.001  & 107 & $<$0.001 \\
T & 1 & 97 & $<$0.001  & 0 & \;\;\;0.83  & 33 & $<$0.001 \\
Z & 1 & 628 & $<$0.001  & 2236 & $<$0.001  & 5 & $<$0.05 \\
C:T & 1 & 823 & $<$0.001  & 7 & $<$0.01  & 40 & $<$0.001 \\
C:Z & 1 & 574 & $<$0.001  & 121 & $<$0.001  & 1 & \;\;\;0.40 \\
T:Z & 1 & 29 & $<$0.001  & 116 & $<$0.001  & 90 & $<$0.001 \\
C:T:Z & 1 & 12 & $<$0.001  & 35 & $<$0.001  & 42 & $<$0.001 \\ \bottomrule
\end{tabular}
\begin{tablenotes}
\item Values are reported for the three error metrics we calculated from the measured point cloud and the respective ground-truth scan: mean of signed error, SD of signed error, and absolute error (AE) after depth shift. 
\item Df = degrees of freedom.
\end{tablenotes}
\end{threeparttable}
\end{table}

\subsection{Experiment 2: Effects of Tissue Type and Zoom}
\label{geometry}
\subsubsection{Depth Accuracy}
\begin{figure}[tbp]
{\includegraphics[width=\columnwidth,trim=0 0 0 .9cm,clip]{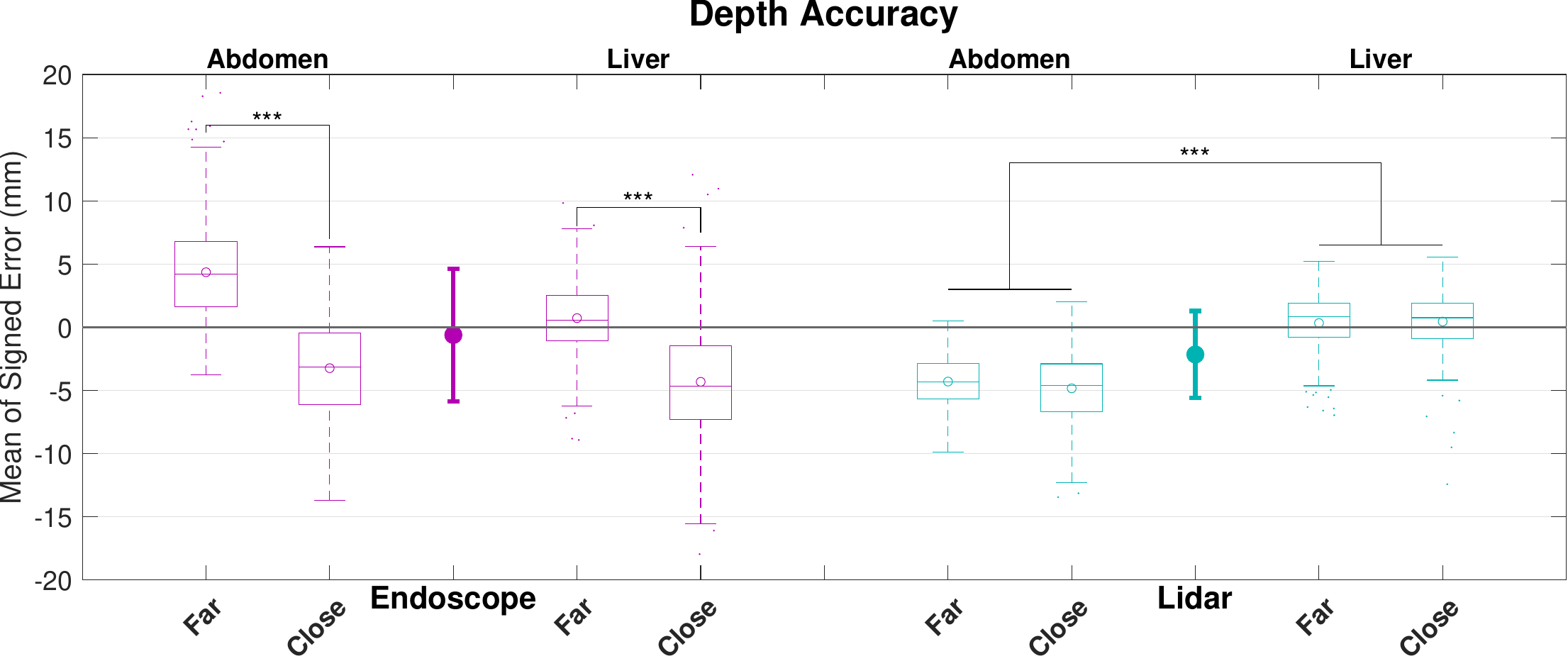}}
\vspace{-.5cm}
\caption{Depth Accuracy for experiment 2: effects of tissue type (Abdomen, Liver) and zoom (Far, Close) on the mean of the signed error.}
\label{Fig:Boxplots_geometry_mean}
\vspace{-.5cm}
\end{figure}

To evaluate Depth Accuracy, we report the \textit{mean of signed error}. This metric represents the ability of the tested 3D cameras to correctly estimate the distance of each point, while its sign shows the direction of any systematic depth shift. 
Fig.~\ref{Fig:averagestd} shows examples of the Depth Accuracy metric color-mapped on point clouds, and the data distributions appear in Fig.~\ref{Fig:Boxplots_geometry_mean} annotated with post-hoc test results. From the results of the ART ANOVA (Table \ref{tab:Models}), the factor that accounts for the largest portion of the variability ($F=823$, $p\leq0.001$) is the interaction between camera and tissue types.

For the \lidar{}, a significant ($p\leq0.0001$) accuracy offset was found between the Abdomen and Liver. Depth error is significantly higher for the Abdomen, with a negative shift (point estimated as more distant) of almost $-5$\,mm on \lidar{} data. This effect was verified for both zoom settings (Far: $p\leq0.0001$; Close: $p\leq0.0001$).
Zoom had no significant effect on \lidar{} accuracy on either tissue type (Abdomen: $p\approx 1$; Liver: $p\approx 1$). However, zoom affected endoscope accuracy, with a significant negative shift of more than $-5$\,mm when moving from Far to Close. This systematic shift occurred for both tissue types (Abdomen: $p\leq0.0001$; Liver: $p\leq0.0001$).

\subsubsection{Time Variability}
\begin{figure}[tbp]
{\includegraphics[width=\columnwidth,trim=0 0 0 .9cm,clip]{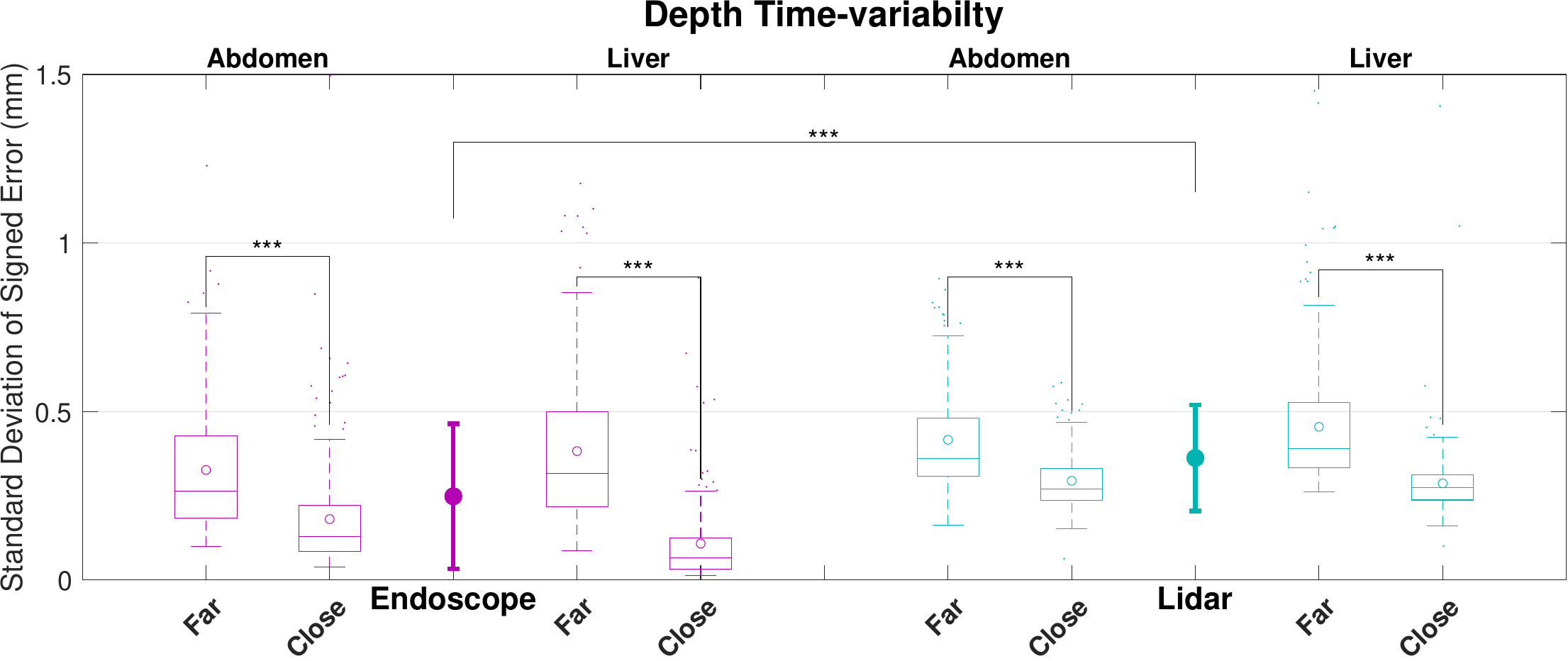}}
\vspace{-.5cm}
\caption{Time Variability for experiment 2: effects of tissue type (Abdomen, Liver) and zoom (Far, Close) on the SD of the signed error.}
\label{Fig:Boxplots_geometry_SD}
\vspace{-.5cm}
\end{figure}

To quantify Time Variability, we report the SD of the signed error (Fig.~\ref{Fig:Boxplots_geometry_SD}); this metric represents the stability with which depth is estimated over time. From the results of the ART ANOVA (Table \ref{tab:Models}), the factor that accounts for the largest portion of the variability ($F=2236$, $p\leq0.0001$) is the zoom setting. Zoom affected Time Variability by showing significantly lower depth SD at Close with respect to Far ($p\leq0.0001$) across all the experimental conditions. Furthermore, the endoscopic data showed a significantly ($p\leq0.0001$) lower Time Variability than \lidar{} data. No statistically significant differences in Time Variability were found between tissue types ($p=0.834$).

\begin{figure}[tbp]
{\includegraphics[width=\columnwidth,trim=0 0 0 .9cm,clip]{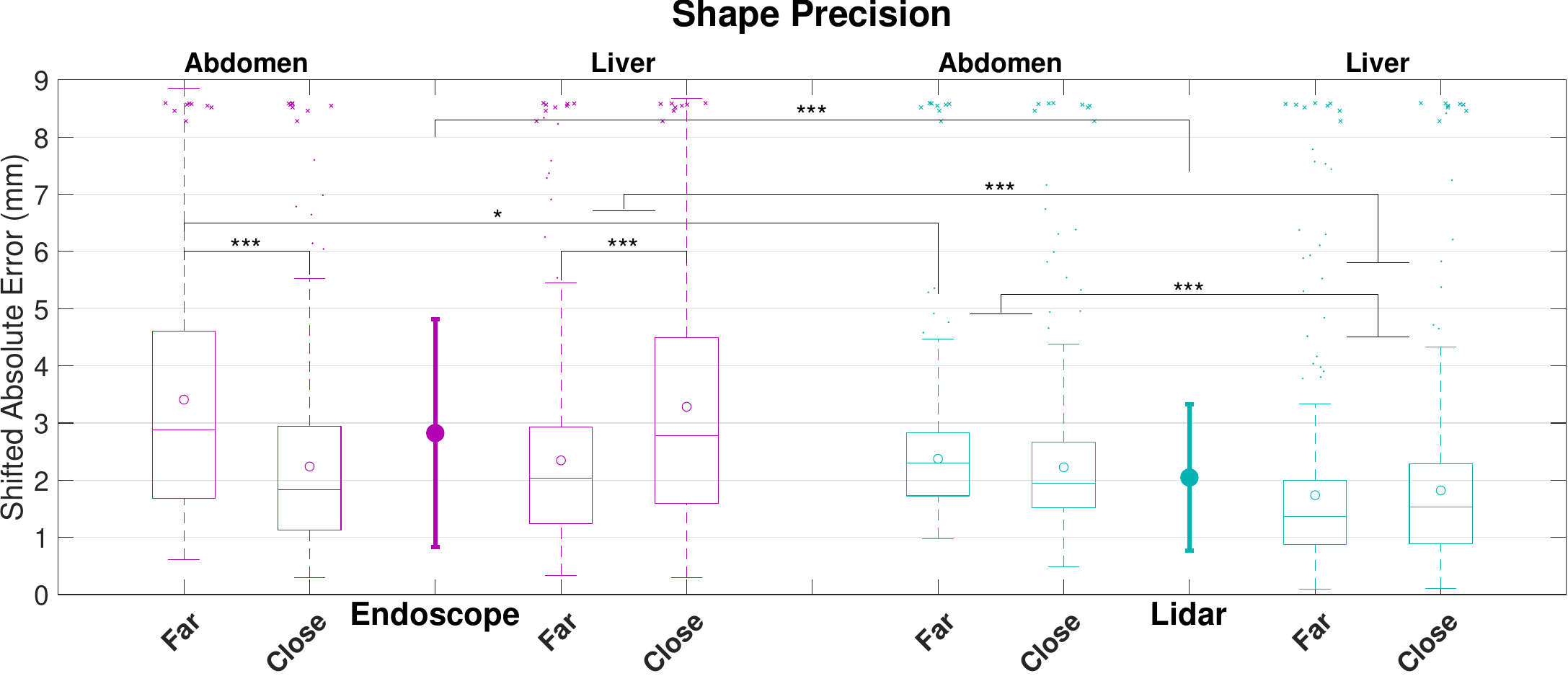}}
\vspace{-.5cm}
\caption{Shape Precision for experiment 2: effects of tissue type (Abdomen, Liver) and zoom (Far, Close) on the shifted AE.}
\label{Fig:Boxplots_geometry_MEA}
\end{figure}

\subsubsection{Shape Precision}
To investigate Shape Precision, we report the shifted AE in Fig.~\ref{Fig:Boxplots_geometry_MEA}; this metric quantifies the local deviation from the ground-truth surface, independent of the absolute depth values measured. From the results of the ART ANOVA (Table \ref{tab:Models}), the factor that accounts for the largest portion of the variability ($F=107$, $p\leq0.0001$) is the camera type. Overall, the \lidar{} shows a significantly ($p\leq0.0001$) better Shape Precision (lower shifted AE) than the endoscope. Specifically, the difference is found for the Liver at both Zoom settings ($p\leq0.0001$) and for the Abdomen at Far ($p\leq0.05$). No difference between the two cameras is found for the Abdomen at Close ($p\approx 1$). Overall, Zoom significantly affected Shape Precision only for the endoscope, with a higher error for the Abdomen at Far ($p\leq0.0001$), and for the Liver at Close ($p\leq0.0001$). Finally, tissue type affected the \lidar{}, with significantly better Shape Precision for the Liver versus the Abdomen at both Zoom settings ($p\leq0.0001$).

\begin{table}[tbp]
\renewcommand{\arraystretch}{1.3}
\centering
\begin{threeparttable}
\caption{Experiment 3 performance comparison via ART ANOVA to investigate the effects of camera type (C), presence of blood (B), illumination (I), and their interactions.}
\label{tab:Models2}
\begin{tabular}{@{}ll|ll|ll|ll@{}}
 \multicolumn{2}{c}{}  &  \multicolumn{2}{c}{\textbf{Depth Accuracy}} & \multicolumn{2}{c}{\textbf{Time Variability}} & \multicolumn{2}{c}{\textbf{Shape Precision}} \\
 \multicolumn{2}{c}{}  &  \multicolumn{2}{c}{Mean} & \multicolumn{2}{c}{SD} & \multicolumn{2}{c}{Shifted AE} \\
\toprule
Model & Df & F & $p$-value & F & $p$-value & F & $p$-value \\ \midrule
C & 1 & 76 & $<$0.001 & 34 & $<$0.001  & 93 & $<$0.001 \\
B & 1 & 4 & $<$0.05  & 1 & \;\;\;0.30  & 21 & $<$0.001 \\
I & 1 & 0 & \;\;\;0.96   & 57 & $<$0.001  & 8 & $<$0.01 \\
C:B & 1 & 6 & $<$0.05  & 3 & \;\;\;0.08 & 11 & $<$0.001 \\
C:I & 1 & 4 & \;\;\;0.06   & 139 & $<$0.001  & 12 & $<$0.001 \\
B:I & 1 & 28 & $<$0.001   & 1 & \;\;\;0.42  & 5 & \;\;\;0.03 \\
C:B:I & 1 & 53 & $<$0.001  & 0 & \;\;\;0.97  & 1 & \;\;\;0.39 \\ \bottomrule
\end{tabular}
\begin{tablenotes}
\item Values are reported for the three error metrics we calculated from the measured point cloud and the respective ground-truth scan: mean of signed error, SD of signed error, and AE after depth shift. 
\item Df = degrees of freedom.
\end{tablenotes}
\end{threeparttable}
\end{table}

\subsection{Experiment 3: Effects of Illumination and Blood}
\label{light}
\subsubsection{Depth Accuracy}
\begin{figure*}[tp]
    \centering
    \includegraphics[width=\textwidth]{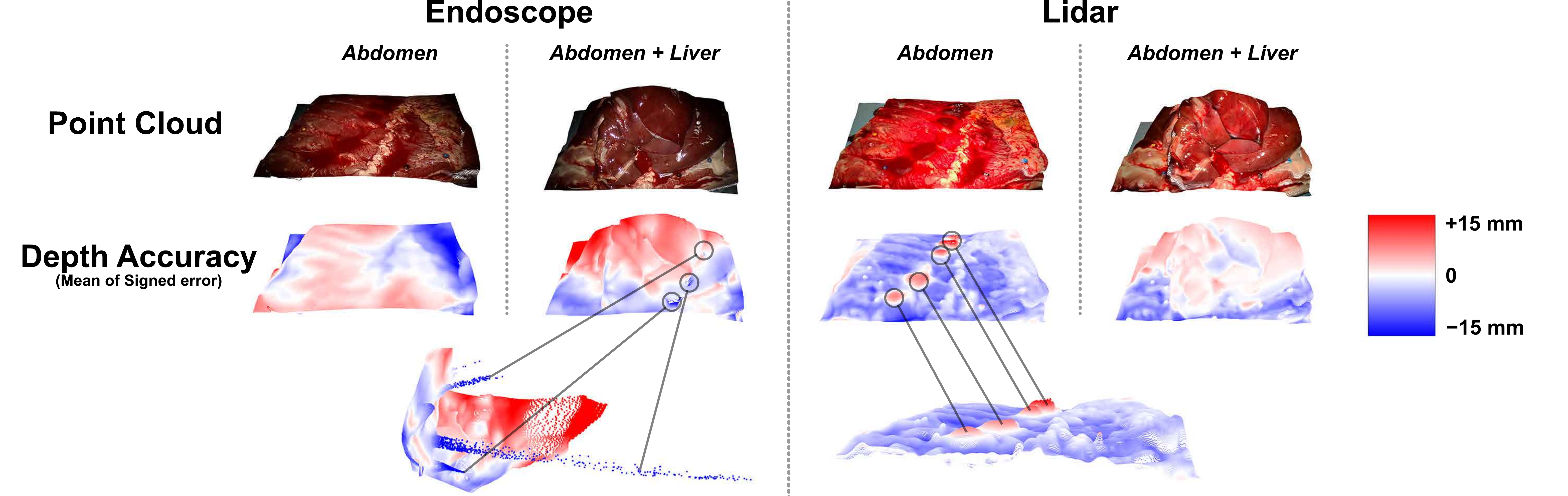}
    \caption{Effects of blood on the tissue surface in experiment 3. The endoscope shows local negative errors (deep holes) due to blood and specular reflections, and the \lidar{} shows local positive errors (round hills) at the pools of blood.}
    \label{Fig:blood}
\end{figure*}

From the results of this experiment's ART ANOVA (Table~\ref{tab:Models2}), the factor that accounts for the largest portion of the variability ($F=76$, $p\leq0.001$) is the camera type. \Lidar{} shows a negative Depth Accuracy shift of almost $-5$\,mm, as similarly reported for the Abdomen in Experiment 2; the lack of content masking in this experiment causes the Abdomen tissue to be visible on all sides of the Liver. Analyzing the local trends, the presence of blood generated a significant positive Depth Accuracy shift (point estimated as closer) for the endoscope at Low light ($p\leq0.001$) and for the \lidar{} at Full illumination ($p\leq0.01$). The opposite effect (point estimated as more distant) is significant for the endoscope ($p\leq0.001$) when Blood and Full illumination are combined (specular light reflections). Examples of local artifacts (point cloud offsets) linked to the presence of blood are highlighted on the signed error fields shown in Fig.~\ref{Fig:blood}.

\begin{figure}[p]
{\includegraphics[width=\columnwidth]{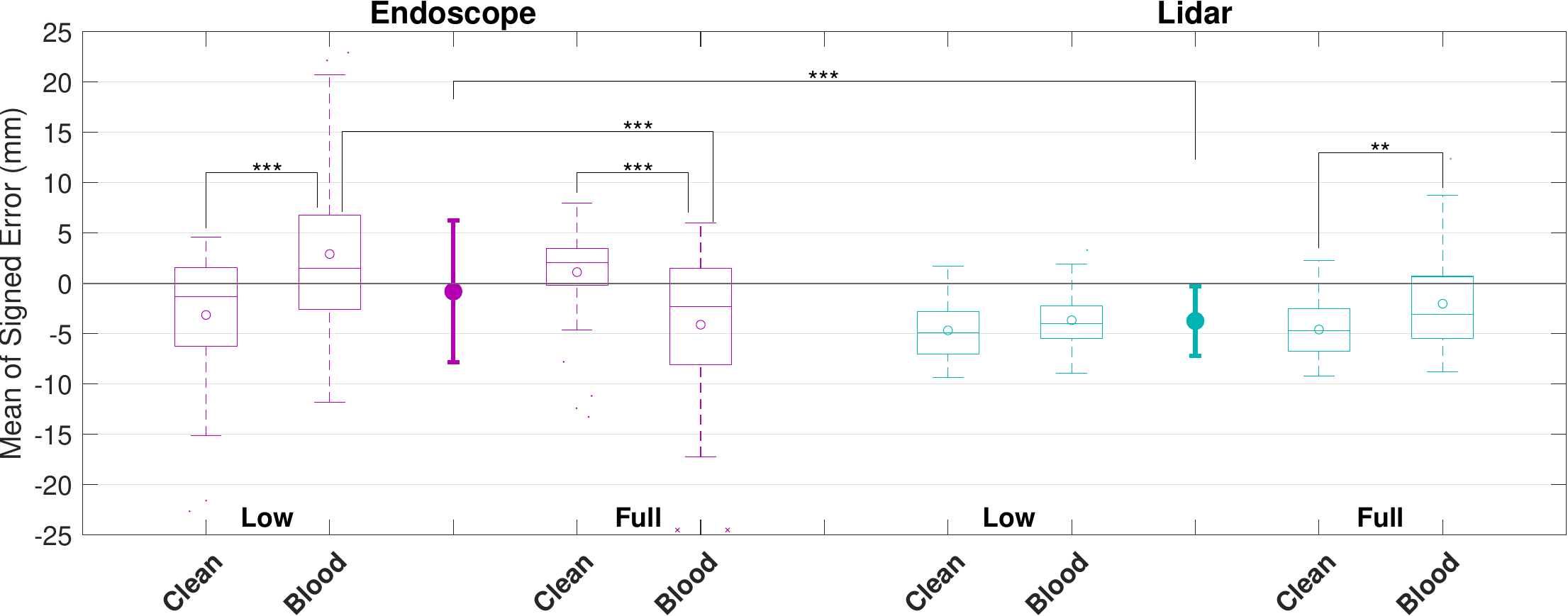}}
\caption{Depth Accuracy for experiment 3: effects of blood (Clean, Blood) and illumination (Low, Full) on the mean of the signed error.
}
\label{Fig:Boxplots_illumination_mean}
\end{figure}

\begin{figure}[p]
{\includegraphics[width=\columnwidth,trim=0 0 0 0cm,clip]{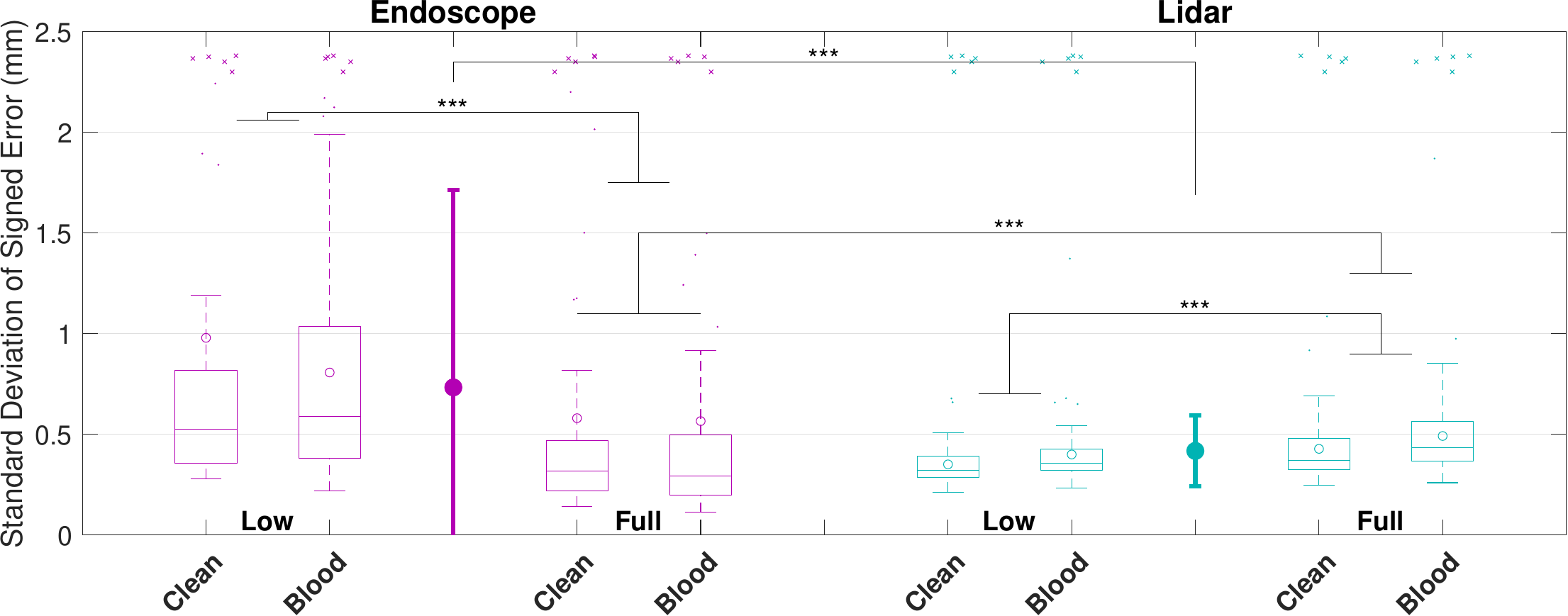}}
\caption{Time Variability for experiment 3: effects of blood (Clean, Blood) and illumination (Low, Full) on the SD of the signed error.}
\label{Fig:Boxplots_illumination_SD}
\end{figure}

\begin{figure}[p]
{\includegraphics[width=\columnwidth]{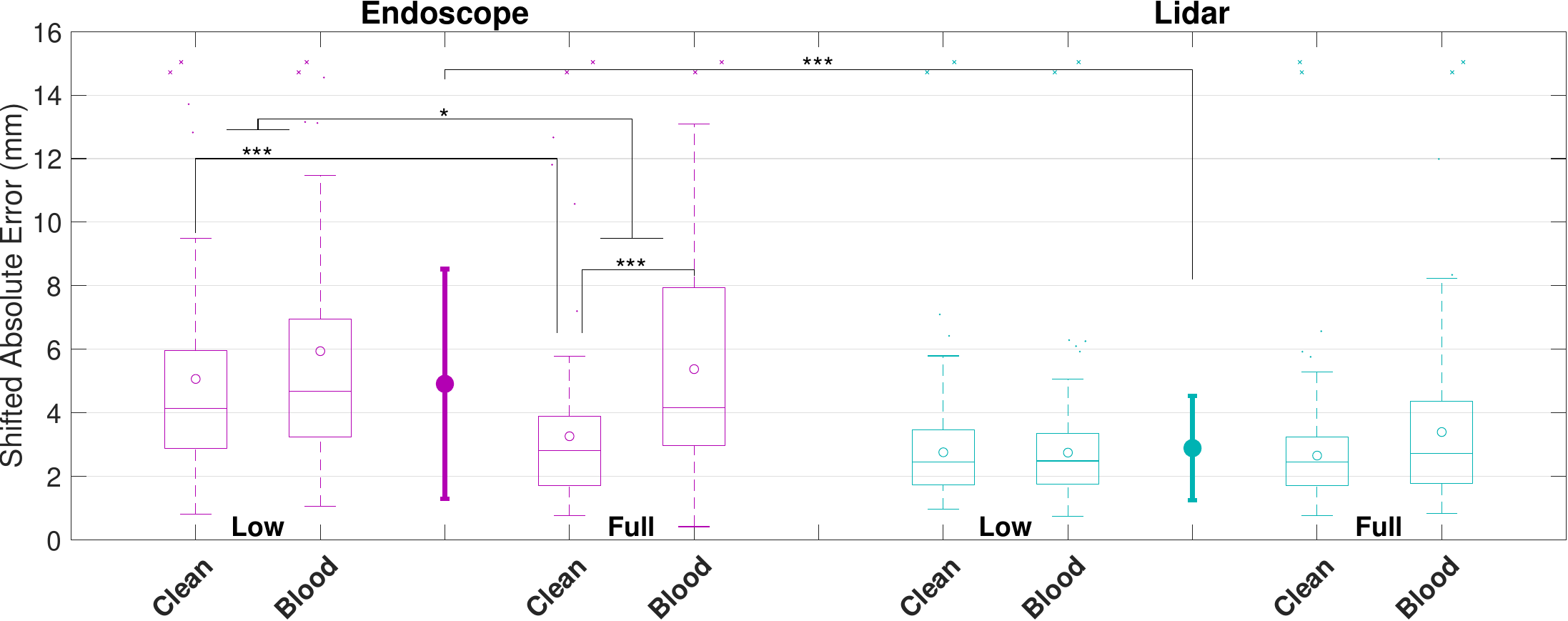}}
\caption{Shape Precision for experiment 3: effects of blood (Clean, Blood) and illumination (Low, Full) on the shifted AE.
}
\label{Fig:Boxplots_illumination_MEA}
\end{figure}

\subsubsection{Time Variability}
The results of the ART ANOVA (Table~\ref{tab:Models2}) show that the factor that accounts for the largest portion of the variability ($F=139$, $p\leq0.001$) is the interaction between camera type and illumination, revealing opposite trends for the two cameras. Time Variability increases for the endoscope ($p\leq0.001$) and decreases ($p\leq0.001$) for the \lidar{} with Low illumination (Fig.~\ref{Fig:Boxplots_illumination_SD}). The \lidar{} shows, in fact, significantly ($p\leq0.001$) lower Time Variability than the endoscope at Low illumination, but significantly higher ($p\leq0.001$) at Full. Overall, the \lidar{} shows significantly ($p\leq0.001$) better Time Variability than the endoscope. Blood has no significant effect on Time Variability ($F=1$, $p=0.30$).

\subsubsection{Shape Precision}
From the results of the ART ANOVA (Table~\ref{tab:Models2}), the factor that accounts for the largest portion of the variability ($F=93$, $p\leq0.001$) is the camera type. Overall, \lidar{} data show significantly ($p\leq0.0001$) better Shape Precision (lower shifted AE) than endoscope data (Fig.~\ref{Fig:Boxplots_illumination_MEA}); this difference is verified at both light settings (Low: $p\leq0.0001$, Full: $p\leq0.0001$). Overall, Blood induces worse Shape Precision (higher shifted AE), though the difference is significant only for the endoscope at Full illumination ($p\leq0.0001$). Low illumination significantly ($p\leq0.0001$) deteriorates the Shape Precision of the endoscope, but not that of the \lidar{} ($p\approx 1$).

%% file: 8_Discussion.tex
In discussing our results, we summarize our major findings, together with a critical assessment of their causes and implications. We then analyze the limitations and potential improvements of our work. 

\paragraph*{Image Processing Time}
We measured a total average processing delay from capture to visualization that is 64.9\% larger for the endoscope ($185$\,ms) than the \lidar{} ($112.2$\,ms) at equal image resolution (nHD, $640\times360$). Additionally, the \lidar{} can generate 3D nHD point clouds at a frequency ($30$\,Hz) that is twice that of the endoscope pipeline. When processing full-resolution frames, the tested stereo-matching algorithm can output 3D point clouds at only $5$\,Hz, which is six times slower than the \lidar{}. Overall, these findings suggest the non-trivial nature of achieving low-latency image processing, especially for endoscopic deep stereo matching. In addition, recent trends toward free-viewpoint 3D visualization through neural rendering, though promising, bring additional computational challenges for real-time 3D perception frameworks. Lower neural inference time together with dedicated hardware capable of low-latency GPU signal processing and rendering could help close this gap.

\paragraph*{Overall 3D Shape Precision}
Both 3D reconstruction techniques showed an average shifted AE in the $1-5$\,mm range, with a grand mean of $2.8\pm2$\,mm for the endoscope and a somewhat lower grand mean of $2.0\pm1.3$\,mm for the \lidar{} in ideal imaging conditions. The respective Shape Precisions degrade greatly to $4.9\pm3.6$\,mm for the endoscope and more moderately to $2.9\pm1.6$\,mm for the \lidar{} in the presence of Blood and Low illumination. These results align with previous findings for stereo matching~\cite{Edwards2021SERV-CT:Reconstruction} and time-of-flight~\cite{Stolyarov2022Sub-millimeterFlight} in endoscopy. They do not yet reach the sub-millimeter accuracy provided by industrial 3D scanners, but such devices do not provide high-frequency continuous real-time point clouds, as \lidar{} does. Furthermore, even though we customized the time-of-flight settings to allow imaging at close range (maximum accuracy in the $10-20$\,cm band), the specific hardware tested here was not engineered for surgery or close-up imaging.

\paragraph*{Interaction with Zoom}
Zoom was found to affect the endoscope's 3D output, with significant depth offsets and precision variations when changing the imaging distance. The same effect was not seen for the \lidar{}, with little or no change in Depth Accuracy and Shape Precision due to Zoom.
This distinction makes sense because stereoscopic depth estimation highly depends on geometric factors, especially the baseline between the left and right cameras and the range of distances for which the camera calibration process was optimized. The tested endoscope has a baseline of only $6$\,mm, which limits performance at farther distances. In addition, the da Vinci Si endoscope has a variable focus. Tiny focus adjustments were made during our experiments when changing the Zoom setting, to keep the images as sharp as possible. While of negligible magnitude in our setting, substantial focus shifts can have an impact on the reliability of the camera calibration. Such a tradeoff is of high relevance when trying to obtain a reliable 3D reconstruction during surgery, where out-of-focus images would impair both the surgeon's perception and the stereo-matching process. Achieving focus-invariant intrinsic calibration is currently an open issue; Khalia et al.\ proposed a solution based on look-up tables~\cite{Kalia2019Marker-lessReality}. New endoscopes promise lower image distortion and invariance to imaging distance via chip-on-tip designs and fixed-focus optics.

\paragraph*{Time Variability}
Overall, both 3D reconstruction techniques showed an average temporal SD in the $0-1$\,mm range with an average SD of $0.25\pm0.2$\,mm for the endoscope and a somewhat higher average SD of $0.36\pm0.16$\,mm for the \lidar{} in ideal imaging conditions. The respective Time Variability degrades greatly to $0.73\pm1.0$\,mm for the endoscope, and more moderately to $0.41\pm0.18$\,mm for the \lidar{} in the presence of Blood and Low illumination. For the endoscope, the high SD is seen at all depths, while the \lidar{}'s Time Variability seems depth-dependent (Fig.~\ref{Fig:averagestd}). Overall, both cameras had lower Time Variability when imaging close-up, and this effect was stronger for the endoscope. 
Such effects can be explained by considering the working principles of the two imaging technologies. Deep stereo matching is based on RGB intensities and, therefore, intrinsically depends on exposure conditions and good matches between views. Barrel distortion and vignetting are common sources of error for endoscopic images. Occlusions, reflections, and complex tissue geometries still represent a challenge for image-based 3D perception. 
In contrast, \lidar{} imaging does not depend on white light exposure, since commercial \lidar{} lasers usually emit in the $800-1300$\,nm spectrum. Depth estimation highly depends on an appropriate match between the laser power, the receiver gain, and the imaging distance. Areas of the workspace that deviate from the configured optimal imaging distance experience larger errors. Adaptive settings and online re-calibration depending on the imaging distance might benefit performance.

\paragraph*{Interaction with Tissue Type}
The biological properties of the observed tissue seem to influence \lidar{} Depth Accuracy. We found a negative offset of approximately $-5$\,mm (point estimated more distant) in the \lidar{}'s Abdomen point clouds. The respective signed error fields (Fig.~\ref{Fig:averagestd} show that portions of the tissue with high fat concentration are estimated with almost zero error (white), while exposed muscle fibers (possibly due to resection) show a negative error (blue). Previous ex-vivo studies have correlated the reflection and absorption coefficients of the infrared spectrum with different porcine samples~\cite{Ding2005Determination1557nm, Bergmann2021ExTissue}. 
Our \lidar{} laser emits at a near-infrared wavelength ($860$\,nm). It is likely that the observed depth shift between muscle, fat, and liver tissue is caused by the interaction between the laser light and the specific tissue density and biological composition. Further investigation might include different tissue thicknesses, multiple laser wavelengths, and pathological tissue samples (e.g., cancerous, ischemic).
The implications of such interactions could inspire the development of a novel imaging technology that combines time-of-flight depth estimation and multi- or hyper-spectral imaging for improved real-time perception. 
Concurrent spectral tissue analysis could enable compensation of tissue-dependent \lidar{} depth offsets. Learning-based solutions seem specifically appealing in this direction. 

\paragraph*{Interaction with Illumination}
The output of the stereo matching showed increased Time Variability and reduced Shape Precision when operating in low-light conditions. Similar results are to be expected during sensor saturation due to extreme illumination. In contrast, \lidar{} showed no performance degradation due to light changes. Curiously, we found a slight decrease in \lidar{} Time Variability with reduced scene illumination, perhaps due to ambient light interference. This experiment underlines the low robustness of vision-based RGB imaging systems in endoscopic surgery. The future development of hardware-based multi-modal sensing and its intrinsic redundancy might facilitate safer and more reliable surgical perception.

\paragraph*{Interaction with Blood}
The presence of blood caused local aberrations for point clouds from both cameras (Fig.~\ref{Fig:blood}). Areas where blood accumulated led to positive shape error for the \lidar{} (surface estimated as closer), and the opposite effect was noticed in endoscope-generated 3D volumes. This last aberration was particularly strong in endoscope imaging when blood was combined with local light reflection (Full illumination). The nature of such artifacts needs to be investigated, especially with more realistic intraoperative lighting and hemorrhagic conditions. 

%% file: 9_Conclusions.tex
In addition to its better precision in 3D shape reconstruction, \lidar{} showed substantial advantages over image-based stereo matching including lower latency, higher capture frequency, and higher robustness to imaging distance and illumination of the scene. 
However, the tested \lidar{}'s accuracy showed a strong dependency on the imaged tissue type, with fat and liver perceived accurately and muscle having a significant depth offset. If it can be thoroughly understood, this shift might open a path for the development of new methods that combine spectral and depth imaging. 

Proper engineering and fine-tuning of a miniaturized time-of-flight sensor for endoscopy seems like a promising path to high-fidelity real-time 3D reconstruction in surgery. We envision the use of \lidar{} with biomimetic infrared markers, as an evolution of the work presented by Decker et al.~\cite{Decker2017BiocompatibleSystem}, to provide real-time intraoperative tissue deformation tracking. Furthermore, we previously proposed the use of multiple \lidar{} cameras attached to the cannulas of a surgical robot~\cite{Caccianiga2022DenseSurgery}. In light of this article's findings, we believe \lidar{} should not immediately replace traditional stereo endoscopy, but rather complement it in a multi-view setting. One or multiple ToF sensors placed at the edges of the abdominal cavity would generate high-speed wide-view 3D imaging and tracking of the anatomy. Concurrently, the main camera (stereo endoscope controlled by the surgeon) would provide an accurate reconstruction of the surgical target from close up. 

Our future work will concentrate on the development of a robust real-time framework for mutual non-rigid registration and fusion of multiple 3D image outputs produced from different perspectives. We believe future computer-integrated surgery will highly benefit from an abundance of intraoperative 3D imaging sources and efficient processing of their outputs. Special attention should be given to intuitive user interfaces and effective 3D visualization strategies.